\documentstyle[psfig,12pt]{article}
\begin{document}
\title{A Non-Gaussian Option Pricing Model with Skew}
\author{Lisa Borland $^{*}$ and Jean-Philippe Bouchaud $^+$ \\ $^{*}$ Evnine-Vaughan Associates, Inc.\\456 Montgomery Street, 
Suite 800, San Francisco, CA 94104,  USA \\ lisa@evafunds.com\\
$^+$ Science \& Finance/Capital Fund Management\\ 6-8 Bd Haussmann, 75009 Paris, France}
\maketitle

\begin{abstract}
Closed form option pricing formulae explaining skew and smile are obtained within a 
parsimonious non-Gaussian framework.
We extend the non-Gaussian option pricing model of L. Borland (Quantitative Finance, {\bf 2}, 415-431, 2002) to include 
volatility-stock
correlations consistent with the leverage effect.
A generalized Black-Scholes partial differential equation for this model
is obtained, together with closed-form approximate solutions for the fair price of a
European call option. In certain limits, the standard Black-Scholes model 
is recovered, as is the Constant Elasticity of Variance (CEV) model of Cox and Ross.  
Alternative methods of solution to that model are thereby also discussed. 
The model parameters are partially fit from empirical observations of the distribution of the underlying.
The option pricing model then predicts European call prices which fit  well to empirical market 
data over several maturities.
\end{abstract}

\section{Introduction}
Over the past three decades, since the seminal works of Black, Scholes and Merton \cite{Black&Scholes,Merton},
the basic ideas of arbitrage-free option pricing have become quite well-known. The Black and Scholes model 
is such a standard  benchmark for calculating option prices, that is used by traders world-wide in spite of 
the fact that the prices it yields do no match ones observed on the market place. 
This difference is clearly a signature of the fact that, after all, the Black-Scholes formula 
is just a model of reality, based on a set of assumptions which
obviously only approximate (often quite remotely) the true dynamics of financial markets. Most importantly,
real markets are incomplete and risk cannot in general be fully hedged away, as it is the case in the 
Black-Scholes world \cite{us-qf}. Nevertheless, prices are quoted in terms of the 
Black-Scholes model, and in order to obtain the market prices from this model, 
it is commonplace to adjust the volatility parameter $\sigma$ which enters the model. 
In other words, traders use a different value of $\sigma$ for 
each value of the option strike price $K$, as well as for each value of the option expiration time $T$. 
This is tantamount to having a continuum of different models (corresponding to each value of $\sigma$) 
for the different options on the very same underlying instrument. A plot of $\sigma$ over strike and maturity 
then yields a surface, often convex and sloping, which is referred to as the smile or the skew, 
or more generally the smile or skew surface. Given a smile surface, one can plug those values of
$\sigma$ into a Black-Scholes model and obtain prices for each strike and time to expiration.

Many attempts have been made to modify or extend the Black-Scholes model in order to accommodate for the skew
observed on option markets.
A large class of those models is based on modeling the skew surface itself. For example, 
so-called local volatility models \cite{Dupire,Derman&Kani} aim to fit the observed skew surface by 
calibrating a function $\sigma_{loc}(S,t)$ (where $S$ is the stock price and $t$ the time) such that it reproduces 
actual market prices for each $K$ and $T$. 
This function $\sigma_{loc}(S,t)$ is then used in conjunction with the Black-Scholes model
to perform other important operations such as the pricing of exotic options, hedging, and so on. The problem is however,
that while $\sigma_{loc}$ by construction allows one to reproduce the set of prices to which it was calibrated, 
it does not contain any 
information about the true dynamics of the underlying asset; therefore it can actually lead to worse hedging strategies 
and erroneous exotic prices than would have been obtained with simply using Black-Scholes without any
attempts to account for the smile \cite{sabr}. Nonetheless, local volatility models have been widely used by many banks 
and trading desks.

L\'evy processes \cite{Eberlein,Levendorski,Cont,jpbouchaud}, 
stochastic volatility models \cite{Heston,Fouque,CGMY,sabr,Pochart,Perello} 
or cumulant expansions around the Black-Scholes case 
\cite{Jarrow,Corrado,Backus,PCB,jpbouchaud} constitute approaches  
which have been successful in capturing some 
of the features of real option prices. For example 
the recent `stochastic alpha, beta, rho' (SABR) model (\cite{sabr},
and see Appendix C) can be well-fit to empirical skew surfaces. It also provides a better model of the dynamics of 
the smile over time.
Nevertheless, in all these models the focus is still on fitting or somehow calibrating the parameters of the model
to match observed option prices. The difference in our current approach will instead be to  introduce a stock price 
model capturing some important features of the empirical distribution of stock returns. 
Our model will be intrinsically more parsimonious 
than stochastic volatility models in the sense that we have just one source of randomness, which also 
allows us to remain within the framework of complete markets. The option pricing methodology 
yields closed form approximate 
formulae based on such a model, thereby predicting option prices rather than fitting parameters to
match  observed market prices. We find good agreement between theoretical and 
traded prices, lending support to 
the possible validity and potential applicability of the model.

\section{Non-Gaussian Stock Price Model with Skew}

The standard Black-Scholes stock price model reads
\begin{equation}
\label{eq:lan}
dS = \mu S dt + \sigma S d \omega
\end{equation}
where $d \omega$ represents a zero mean Brownian random noise correlated in time $t$ as
\begin{equation}
\langle d \omega(t) d\omega(t') \rangle _F = dt dt'\delta(t-t'),
\end{equation}
Here, $\mu$ represents the rate of return  and $\sigma$ the volatility of log stock returns.
This model implies that stock returns follow a log-normal distribution, which is only a very rough approximate
description of the actual situation. In fact, real returns have strong power-law tails which are not at all accounted
for within the standard theory. Furthermore, there is a skew in the distribution such that there is a  
higher probability of large negative returns than large positive ones. Both the tails 
and the skew depend on the time over which the returns  are calculated. 
In general, it is observed that the power-law statistics of the distributions are very stable, 
exhibiting tails decaying as $-3$ in the cumulative distribution for returns taken over time-scales
ranging from minutes to weeks, only slowly converging to Gaussian statistics 
for very long time-scales \cite{stanley,jpbouchaud}. Similarly, the skew of the distribution 
varies with the time-scale of returns such that it is largest for intermediate time-scales \cite{jpbouchaud}.

The deviations of the statistics of real returns to those of the log-normal model of Eq. (\ref{eq:lan})
become particularly important when it comes to calculating the fair price of options on the underlying stock. 
%An option is a financial instrument which  gives the owner the right to execute some agreed upon transaction 
%at some time in the future.  
For example, the simplest `European' call option, which is the right to buy the stock $S$ at 
the strike price $K$ at a given expiration time $T$. The call will expire worthless if 
$S(T) < K$, and profitably otherwise. The fair
price of the call thus depends on the probability that the stock price $S(T)$ exceeds $K$, and if 
one uses the wrong statistics in the model then the theoretical price will differ quite a bit from 
the empirically traded price. (Interestingly enough, market players intuitively adjust the price of options 
to be much more consistent with the true statistics of stock returns \cite{PCB}, even though the log-normal 
Black-Scholes model is widely used by traders themselves to get an estimate of what the price should be.) 

In this paper, we develop a model of the underlying stock which is consistent with both fat-tails and skewness 
in the returns distribution. We then derive option prices for this model, obtaining (approximate) closed-form solutions 
for European call options. 

Our model bases on the non-Gaussian model \cite{qf_borland,prl_borland}, where it was proposed that 
the fluctuations driving stock returns could be modeled by a `statistical feedback' process \cite{pre_borland}, 
namely
\begin{equation}
\label{eq:lanq}
dS = \mu S dt + \sigma S d \Omega
\end{equation}
where
\begin{equation}
\label{eq:Omega}
d\Omega = P(\Omega)^{\frac{1-q}{2}} d\omega.
\end{equation}
In this equation, $P$ corresponds to the probability distribution of $\Omega$, which simultaneously evolves
according to the corresponding nonlinear Fokker-Planck equation \cite{Tsallis&Bukman}
\begin{equation}
\label{eq:nlfp}
\frac{\partial P}{\partial t}  = \frac{1}{2} \frac{\partial P^{2-q}}{\partial \Omega^2}.
\end{equation}
The index $q$ will be taken $3 \ge q \ge 1$. In that case, Eq. (\ref{eq:nlfp}) is known also as the fast diffusion 
equation \cite{Chasseigne&Vasquez}. Note also that in the above equation, the 
time $t$ must be a-dimensional. In the following, the unit of time has 
been chosen to be one year. Correspondingly, $\sigma$ is also a-dimensional.   

Equation (\ref{eq:nlfp}) can be solved exactly, leading, when the 
initial condition on $P$ is a $P(\Omega,t=0)=\delta(\Omega)$, to a Student-t (or
Tsallis \cite{andre}) distribution: 
\begin{equation}
\label{eq:ptsallis}
P = \frac{1}{Z(t)}\left(1 + (q-1) \beta(t) \Omega^2(t)\right)^{-\frac{1}{q-1}}
\end{equation} 
with 
\begin{equation}
\label{eq:beta}
\beta(t) = c_q^{\frac{1-q}{3-q}}\left((2-q) (3-q) \, t\right)^{-\frac{2}{3-q}}
\end{equation} 
and
\begin{equation}
\label{eq:Z}
Z(t) = \left( (2-q) (3-q) \, c_q t \right)^{\frac{1}{3-q}} 
\end{equation} 
where the $q$-dependent constant $c_q$ is given by
\begin{equation}
c_q = \left[\int_{- \infty}^{\infty} ( 1 + (q-1)u^2)^{-\frac{1}{q-1}} \, du\right]^2 \equiv{\frac{\pi}{q-1}} \frac{\Gamma^2(\frac{1}{q-1}-
\frac{1}{2})}{\Gamma^2(\frac{1}{q-1})}.
\end{equation}

Eq. (\ref{eq:ptsallis}) recovers a Gaussian in the limit 
$q \rightarrow 1$ while exhibiting power law tails for all $q > 1$. 

The statistical feedback term $P$ can also be seen as a price-dependent 
volatility that captures the market sentiment.
Intuitively, this means that if the market players observe {\it unusually large} 
deviations of $\Omega$ (which - after removing a noise induced drift 
term which could equivalently have been
absorbed in the dynamics of Eq. (\ref{eq:lanq}) \cite{qf_borland} -  
is essentially
equal to the detrended and normalized log stock price) from its mean, 
then the effective 
volatility will be high because in such cases $P(\Omega)$ is small, 
and the exponent 
$q$ is larger than unity.
Conversely, traders will react more moderately if $\Omega$ is close 
to its more typical values.
As a result, the model exhibits intermittent behaviour consistent 
with that observed in the effective volatility 
of markets -- but see the discussion below.

Option pricing based on the price dynamics elucidated above was solved in \cite{qf_borland}, and it was seen that 
those prices agreed very well with traded prices for instruments which have a symmetric underlying distribution, 
such as certain foreign exchange currency markets. However, the question of skew was not discussed in 
that paper and is instead the topic of the current work. 
We extend the stock price model to include an effective volatility that
is consistent with the leverage correlation effect, namely 
\begin{equation}
\label{eq:Slangevin}
dS = \mu S dt + \sigma  S_0^{1-\alpha} S^{\alpha} d \Omega
\end{equation}
with $\Omega$ evolving according to Eq. (\ref{eq:Omega}). The parameter $\alpha$
introduces an asymmetric skew into the distribution of log stock returns. More precisely, when 
$\alpha < 1$, the {\it relative} volatility can be seen to increase when $S$ decreases, and 
vice-versa, an effect known as the leverage correlation (see \cite{leverage}). 
For $\alpha = q = 1$ the standard Black-Scholes model is recovered. For $\alpha = 1$ but
$q > 1$ the model reduces to that discussed in \cite{qf_borland}, while for $q=1$
but general $\alpha$ it becomes the constant elasticity of variance (CEV) model of Cox and Ross \cite{Cox&Ross}.  

There are two possible interpretations to model  Eq. (\ref{eq:Slangevin}), and some limitations which we 
elucidate here. In the context of option pricing,
the relevant question concerns the forward probability, estimated from now ($t=0$), with the current price 
$S_0$ corresponding to the reference price around which deviations are measured. In this case, the fact that 
$t=0$ and $S=S_0$ (or $\Omega=0$) play a special role makes perfect sense. If, on the other hand, one wants 
to interpret Eq. (\ref{eq:Slangevin}) as a model for the real returns, then the choice of $S_0=S(t=0)$ as the
reference price is somewhat arbitrary and therefore problematic. Still, this model produces returns which have 
many features consistent with real stock returns. This is illustrated in Figure 1 where we have plotted 
a time-series of simulated returns as well as, in Figure 2,  a plot of the distribution of returns between times $t$ and 
$t+\tau$, for different time lags $\tau$, averaged over all possible starting times $t$.  Clearly, the model 
reproduces volatility clustering. Also, the returns distribution exhibits fat tails becoming Gaussian over 
larger time scales. (Note that this is not in contradiction with the fact that returns counted from $t=0$ 
have a Student-t distribution for all times $t$).  The skew in the distribution is also apparent.
For visual comparison, we  show the same data for the SP500. The qualitative behaviour of the two data sets
is very similar. However, in order to have a consistent model of real stock returns, one should allow the 
reference price to be itself time dependent. A possibility we are presently investigating is to write 
the statistical feedback term as $P(\Omega - \bar{\Omega})^{(1-q)/2}$, where $\bar{\Omega}$ 
is a moving average of past values of $\Omega$. (The current model corresponds to $\bar{\Omega} = 0$.) 
One can also easily alter the super-diffusive behaviour (as given by Eq(\ref{eq:beta})) of the 
distribution if a mean reversion of the fluctuations is included. Both of these features make the model
a more realistic one of real returns (where distributions are non-Gaussian, yet volatilities are normally diffusive 
across all time-scales). Incidentally, this super-diffusive volatility scaling can alternatively be modified by a 
redefinition of time (see also the discussion later in this paper). 

%Furthermore, we are deriving option pricing formulae within this framework. 

One point necessary to comment on regarding Eq. (\ref{eq:Slangevin}), is that depending on the value of $\alpha$,
stock prices may go negative. We do not allow this to happen: rather, we absorb the stock if the price hits zero.
Basically, this means that the  company has gone bankrupt, or that the stock is trading at such low prices
(penny stocks) that it is practically worthless and we so not consider it part of the trading universe.
It is interesting to look at the probability of bankruptcy $\cal P$ as a function of $\alpha$. This is shown in Figure 3. 
An analytical treatment \cite{usunpub}, expected to be valid for $\eta=\sigma (1-\alpha) \ll 1$, predicts that
\begin{equation}
\label{eq:calP}
{\cal P}= a T^{\frac{1}{q-1}} \eta^{ \frac{3-q}{q-1}}
\end{equation}
 (where $a$ is a computable constant), in good agreement with our data for small 
${\cal P}$.  
In fact, this point suggests a connection with credit risk markets, 
where the probability of default 
is an important quantity. One could imagine using a variant of  our model to model default risk; conversely, one could 
imagine using data of probability of default to help choose the correct value of $\alpha$ of a particular stock.
The consequence of this default probability for option pricing will be discussed in Section 5.

\begin{figure}[t]
\psfig{file=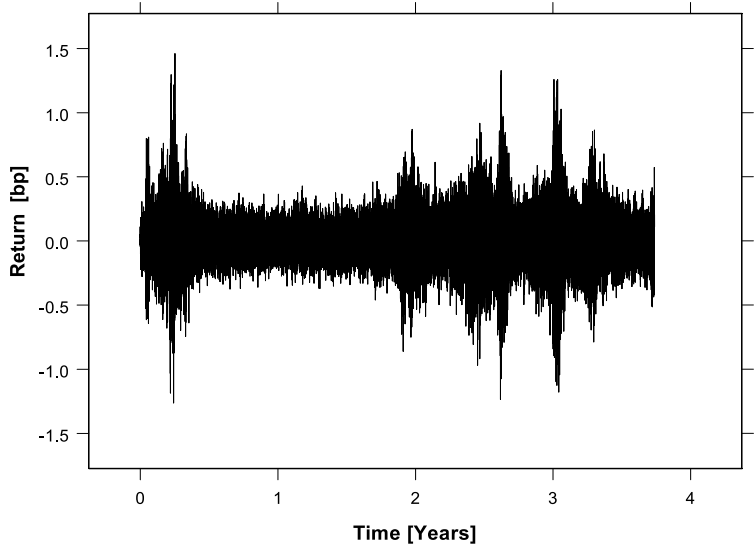,width=4.5 in}
\caption{\footnotesize
A typical path of the non-Gaussian model with $q=1.5$ is shown. Here $\alpha=-3$ and $\sigma=30 \%$.
}
\end{figure}

\begin{figure}[t]
\psfig{file=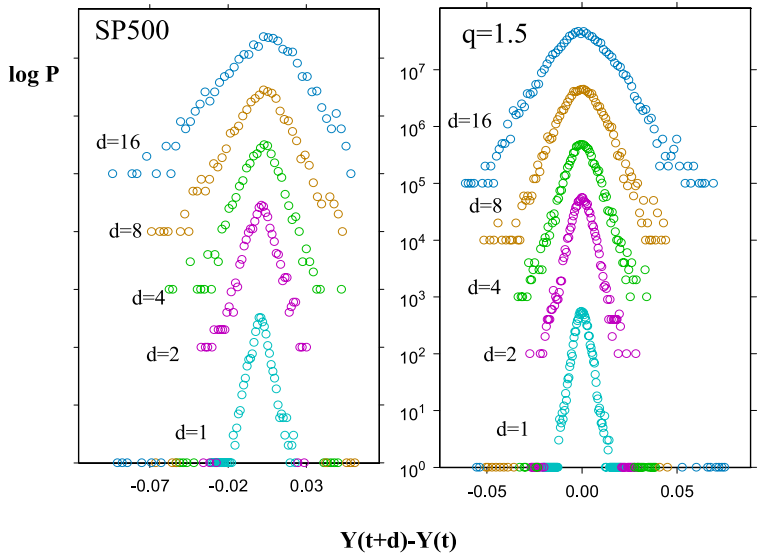,width= 4.5 in}
\caption{\footnotesize
This plot exhibits the distribution of log returns ($Y=\ln S$) for the S\&P500 over different time lags $\tau$ 
ranging from 1 day to 16,
together with corresponding histograms obtained by our model with $q=1.5$ and $\alpha =1$. 
}
\end{figure}

\begin{figure}[t]
\psfig{file=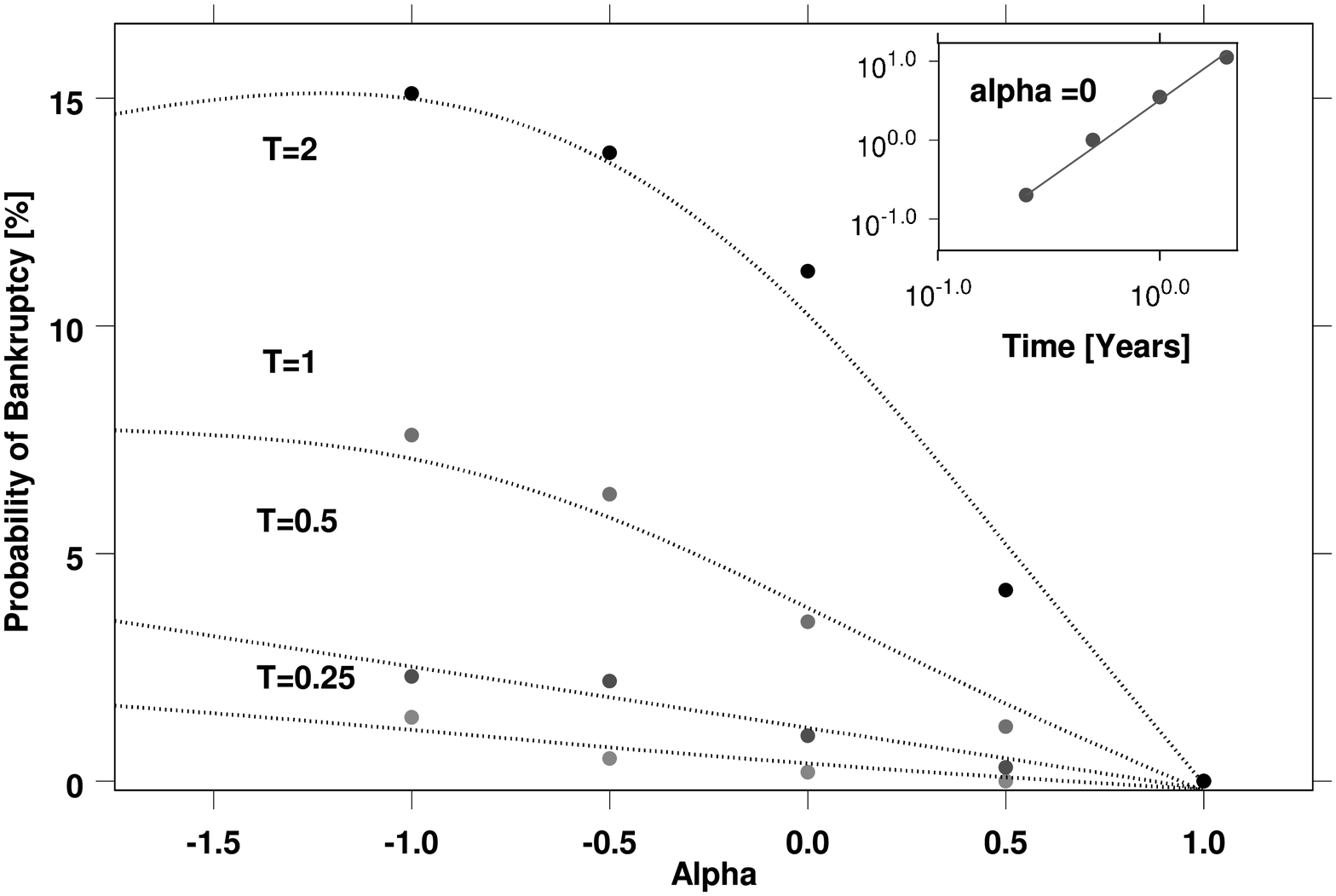,width= 2.5 in}
\psfig{file=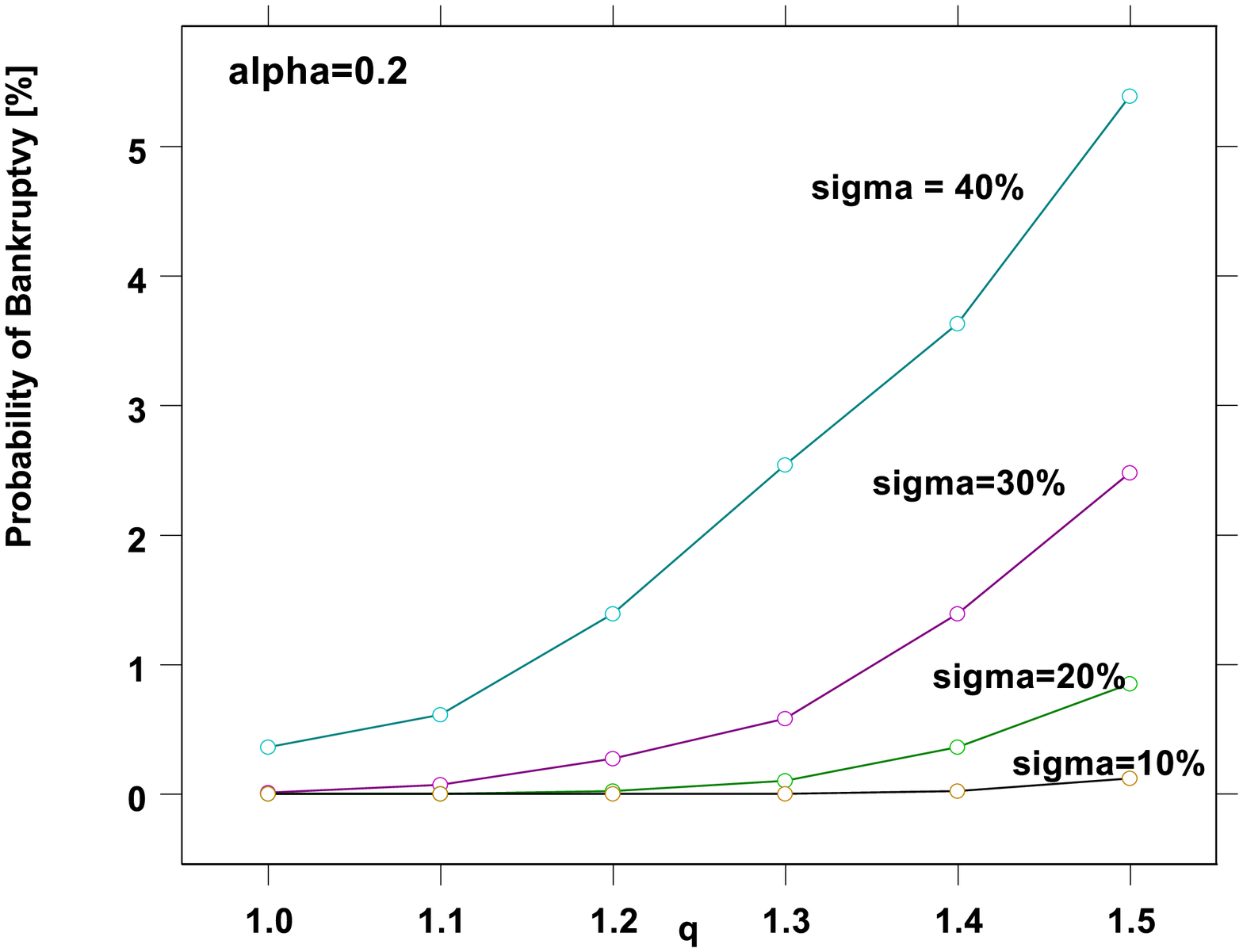,width=2.5 in}
\caption{\footnotesize
a) The probability of bankruptcy (default) $\cal P$ as a function of $\alpha$ for our model with $q=1.5$, 
for times ranging from $0.25$ to $2$ years. The inset shows the probability of bankruptcy as a funtion of time
for a stock with $\alpha=0$, compared to our prediction that ${\cal P}
\propto T^2$ for $q=1.5$.
These results are based on Monte-Carlo simulations with 
$\sigma= 30 \%$, $r = 6 \%$ and $S(0) = \$50$.
b) The probability of bankruptcy is sensitive to the values of $q$ and $\sigma$ as shown here,
for $S(0) = \$50$,  $\alpha=0.2$, $r=4\%$ and $T=1$. 
For $T<1$ these probabilities are quickly depressed according to Eq(\ref{eq:calP}).
}
\end{figure}

\section{Fair Price of Options}

Our model contains only one source of randomness, $\omega$, which is a standard Brownian noise. 
Therefore, the market is complete and usual hedging arguments are valid, as was shown in \cite{qf_borland} for the case
$\alpha = 1$. It is however trivial to see that these results are also valid for arbitrary $\alpha$.
This means that we can immediately adopt many of the arguments usually associated with the Black-Scholes
log-normal world - even though we here have a process with non-Gaussian statistics. In particular, it is possible 
to define a unique equivalent martingale measure $Q$ to the process, related to the original measure $F$ 
via the Radon-Nikodym derivative \cite{mathfinance1,mathfinance2},  such that the discounted stock price 
  $G = S e ^{-rt}$ is a martingale, where $r$ corresponds to the risk-free rate.
 With respect to  $Q$ the process of Eq. (\ref{eq:Slangevin})
reads
\begin{equation}
\label{eq:Smartingale}
dS = r S dt + \sigma S_0^{1-\alpha} S^{\alpha} d \Omega
\end{equation}
with
\begin{equation}
\label{eq:domegaz}
d \Omega = P^{\frac{1-q}{2}} d z
\end{equation}
where the noise $z$ satisfies
\begin{equation}
\langle d z(t') d z (t) \rangle _Q =  dt dt' \delta (t'-t)
\end{equation}
Essentially then, we have replaced $\mu = r$ just as in the standard theory.
Our task now is to solve for the option prices based on the model Eq. (\ref{eq:Smartingale}).

The dynamics for $G$ read
\begin{eqnarray}
dG & = & \sigma S_0^{1-\alpha} G^{\alpha} e^{(\alpha -1)rt} d \Omega\\
& = & \sigma S_0^{1-\alpha} G^{\alpha} P^{\frac{1-q}{2}} e^{(\alpha-1)rt} dz
\end{eqnarray}
We redefine time by introducing the new variable $\hat{t}$
such that 
\begin{equation}
\label{eq:that}
\hat{t} = \frac{e^{2(\alpha-1)rt} - 1}{2(\alpha-1)r}.
\end{equation}
Note that $\hat t \to t$ when $(\alpha-1)rt \to 0$. With respect to $\hat{t}$, 
the Brownian noise $z$ satisfies
\begin{equation}
\langle dz(\hat{t}') d z (\hat{t}) \rangle _Q  =  e^{2 (\alpha-1)rt} \langle dz(t') dz(t)  \rangle _Q
\end{equation}
resulting in
\begin{eqnarray}
\label{eq:gthat}
dG &= &\sigma S_0^{1-\alpha}  G^{\alpha} d\Omega\\
d \Omega(\hat{t}) &=& P(\Omega(\hat{t}))^{\frac{(1-q)}{2}} dz(\hat{t})
\end{eqnarray}
Next, make the variable transformation
\begin{equation}
x = \frac{(G/S_0)^{1-\alpha}-1}{1-\alpha}
\end{equation}
(which simply becomes $x = \ln G/S_0$ for $\alpha =1$), so that
\begin{equation}
\label{eq:xhat}
dx(\hat{t}) = -\frac{\alpha}{2}\sigma^2 \frac{P^{1-q}}{1 + (1-\alpha)x(\hat{t})} d\hat{t} + \sigma d\Omega(\hat{t})
\end{equation}
where the initial conditions read
\begin{equation}
x_0 \equiv x(0) = 0\\
\end{equation}
and where, at the real final time $T$, we have 
\begin{equation}
\label{eq:ST}
S(T) = e^{rT}\left( 1 + (1-\alpha) x(\hat{T}) \right)^{\frac{1}{1-\alpha}}
\end{equation}

Given a representation of the stock price with respect to the risk-neutral world in which the discounted price is 
a martingale, the fair value of a derivative of the underlying stock can be calculated as the expectation of the payoff
of the option. In this paper we shall focus on this approach. However we want to point out that one can equivalently 
obtain the option price by solving the generalized Black-Scholes partial differential equation for the current problem,
which can readily be written down following the lines of \cite{qf_borland}. It reads 
\begin{equation}
\label{eq:bsgen}
\frac{df}{dt}  + rS \frac{df}{dS}  + \frac{1}{2} \frac{d^2f}{dS^2} \sigma^2 
S_0^{2(1-\alpha)}S^{2\alpha} P_q^{1-q} 
= rf  
\end{equation}
where $P_q$  evolves according to Eq. (\ref{eq:ptsallis}).
In the limit $q \rightarrow 1$ and $\alpha \rightarrow 1$, we recover the standard Black-Scholes
differential equation. In the limit $\alpha \rightarrow 1$ we recover the case of \cite{qf_borland}.

In the following 
we proceed to study closed form option pricing formulas obtained via expectations.
 The most immediate and simplest
path is to utilize our knowledge of the statistical properties of the random variable
$\Omega(\hat{t})$. By invoking approximations valid if $\sigma^2 T \ll  1$, which is certainly true for stock 
returns for reasonable maturities, we can obtain closed form solutions for European calls, much along the lines
followed in \cite{qf_borland}. However, we allude in Appendix B to a more complicated path which is 
based on mapping the process onto a higher dimensional process.

\section{Solutions via a Generalized Feynman-Kac Approach}

If $\sigma^2 T \ll 1$, which is valid even for relatively high volatility 
stocks, (e.g. 
typical volatilities of $10 \%$  
to $30 \%$  yield  $\sigma^2 T$  values of .01 to .09 for $T=1$ year), 
then we can 
insert the approximation $x(\hat{t}) \approx \sigma \Omega(\hat{t})$
into the right hand side of Eq. (\ref{eq:xhat}) yielding,
\begin{equation}
\label{eq:sigsmall}
x(\hat{T}) = \sigma \Omega(\hat{T}) - \frac{\alpha}{2} \sigma^2 
\int_0^{\hat{T}} \frac{P(\Omega(\hat{t}))^{1-q}}{1
+ (1-\alpha)\sigma \Omega(\hat{t})} d\hat{t},
\end{equation}
plus order $\sigma^4$ corrections. For  $\alpha =1$ and general $q$, the problem reduces to that 
discussed in \cite{qf_borland}, which  we shall revisit with a slightly different evaluation technique below.

The integral in Eq. (\ref{eq:sigsmall}) contains terms of the type 
\begin{equation}
\label{eq:fint}
\int_0^T F(\Omega(t))dt,
\end{equation}
 in other words integrals of a function $F$
of the random path $\Omega(t)$, conditioned on ending at a particular value $\Omega(T)=\Omega_T$. One way to evaluate 
the integral, which is the approach taken in \cite{qf_borland}, 
is to invoke the following (which is only valid approximately \cite{corrqf}): Replace the 
actual paths $\Omega(t)$ with other paths, such that the ensemble average of the two sets of paths are the same.
Such  paths can be obtained by exploiting the scaling properties of the time-dependent 
probability distribution $P(\Omega(t))$. We know that the variable $\Omega(t)$ at time $t$ is  distributed according to
Eq. (\ref{eq:ptsallis}) above. Due to scaling, we know that we can map $\Omega(t)$ onto the terminal value $\Omega_T$ 
in the following way:
\begin{equation}
\label{eq:approx}
\Omega(t) = \sqrt{\frac{\beta(T)}{\beta(t)}} \Omega_T
\end{equation}
where $\Omega(T) $ is distributed according to Eq. (\ref{eq:ptsallis}) evaluated at time $T$. 
Therefore,
as we illustrate in  Appendix A, the ensemble statistics of this replacement path and the original path 
are equivalent. This approach was shown in \cite{qf_borland} to be quite precise numerically. However, as we discuss 
next, there is another more exact approximation which can be used to evaluate Eq. (\ref{eq:fint}). 
This entails solving a Feynman-Kac equation {\it exactly} up to first order in $\sigma^2$.

Define the quantity
\begin{equation}
W(\Omega_T,T) =\sum_{i\mid \Omega_T} w_i \exp\{\epsilon \int_0^T F(\Omega(t)) dt\}
\end{equation}
which is the expectation calculated as the sum over all paths $i$ ending at $\Omega_T$ at time $T$, 
of the exponential of quantity which we wish to calculate. The coefficients $w_i$ denote
the weights of the paths, and $\epsilon$ is a small parameter.
The function $F$ can be a general one of $\Omega(t)$, but in our current problem we shall be only interested in 
polynomial expansion up to the second order in $\Omega$. We can write
\begin{equation}
W(\Omega_T,T)=\sum_{i\mid \Omega_T} w_i \exp\{\epsilon \int_0^T (\sum_{j=1}^2 f_j(t) \Omega(t)^j ) dt\}
\end{equation}
In the case where the paths evolve according to the nonlinear Fokker-Planck equation, one can establish a generalized
 Feynman-Kac equation for $W(\Omega_T,T)$ that reads:
\begin{equation}
\label{eq:FeynmanKac}
\frac{\partial W}{\partial T} = \frac{1}{2} \frac{\partial^2 W_0^{1-q} W}{\partial \Omega^2} + 
\epsilon \sum_j f_j(T)\Omega_T^j
\end{equation}
where $W_0=P_q(\Omega_T,T)$ represents the solution to the fast diffusion problem, corresponding to $\epsilon = 0$.
It is relatively straightforward to insert the Ansatz
\begin{equation}
\label{eq:fkAnsatz}
W = W_0 \left[1 + \epsilon (g_0(T) + g_1(T) \Omega_T + g_2(T) \Omega_T^2)\right]
\end{equation} 
into Eq. (\ref{eq:FeynmanKac}) and obtain an exact set of differential equations for the time dependent 
coefficients $g_j$, given in Appendix A (Eq. (\ref{eq:fkcoeffsf}) - Eq. (\ref{eq:fkcoeffsg})). 

First, let us illustrate this approach on the example of $\alpha =1$ which corresponds 
to  the problem discussed in \cite{qf_borland}. The expression for $S(T)$ becomes 
(from Eq. ({\ref{eq:ST}) and Eq. (\ref{eq:sigsmall}) in the limit $\alpha \rightarrow 1$)
\begin{equation}
\label{eq:STalpha1}
S(T) = S(0)\exp\{rT +\sigma \Omega_T -\frac{\sigma^2}{2} 
\int_0^T Z(t)^{1-q} (1+(q-1) \beta(t) \Omega(t)^2) dt \} 
\end{equation}
Using the Feynman-Kac formula to evaluate
\begin{equation}
\label{eq:oldpaperterm}
\int_0^T Z(t)^{1-q}\beta(t) \Omega(t)^2 dt
\end{equation}
 (see Appendix A, Eq.(\ref{eq:fkomega2}) with 
$\epsilon = \sigma^2/2$ and $h_2 = Z^{1-q}\beta$), we 
 obtain an expression of the form: 
\begin{equation}
\label{eq:oldpaperterm1}
\langle \int_0^T Z(t)^{1-q}\beta(t) \Omega(t)^2 dt \rangle = 
g_0(T) + g_2(T) \Omega_T^2
\end{equation}
($g_1$ is zero by symmetry in this case). For the stock price, we 
thus obtain
\begin{equation}
\label{eq:stoldfk}
S(T) = S(0)\exp\{rT +\sigma \Omega_T -\frac{\sigma^2}{2} \left[\gamma(T)\frac{3-q}{2} + (q-1)(g_0(T) + g_2(T)\Omega_T^2)\right] \} 
\end{equation}
with $\gamma$ and $g_j$ are given by Eq. (\ref{eq:fkf}) - 
Eq. (\ref{eq:gammaT}). The above expression is exact to order $\sigma^2$;
to that order, the path to path fluctuations of 
$\int_0^T Z(t)^{1-q}\beta(t) \Omega(t)^2 dt$ conditioned on a given value
of $\Omega_T$ can be neglected.  

The coefficients $g_j$ found 
here are slightly different than those in \cite{qf_borland}, 
where instead of 
Eq. (\ref{eq:oldpaperterm}), the approximation Eq. (\ref{eq:approx}) was used. However,  the results are 
numerically very close, so that in practice either evaluation method can be used; both give option price 
which match well to Monte-Carlo simulations. 
In fact, why this is so becomes clear from the plots shown in Appendix A, 
where Monte-Carlo simulations of the quantity Eq. (\ref{eq:oldpaperterm}) 
are compared with the Feynman-Kac 
approximation and the one of \cite{qf_borland}.
In the general case where $\alpha$ is not restricted to unity, 
the challenge now is to evaluate Eq. (\ref{eq:sigsmall}) 
which can be rewritten as: 
\begin{equation}
\label{eq:xhatT}
x(\hat{T}) =  \sigma \Omega_{\hat{T}} - \frac{\alpha \sigma^2}{2} \int_0^{\hat{T}}
Z(t)^{1-q}\frac{1 + (q-1)\beta(t)\Omega(t)^2}{1 + (1-\alpha)\sigma \Omega(t)} dt 
\end{equation}
We shall focus our discussion on the last term.
A solution can be found using a two-step approach: First of all, we make the assumption that the expression, 
when integrated to abitrary time $u$, can be well-described by a Pad\'e approximation, namely a ratio of polynomials, which has the
correct asymptotic behaviour for $\Omega \to 0$ and $\Omega \to \infty$, resulting in 
\begin{equation}
\label{eq:xhatTPade}
x(\hat T) =  \sigma \Omega_{\hat T} - \frac{\alpha \sigma^2}{2} 
\frac{A(\hat T) + B(\hat T)\Omega_{\hat T} + C(\hat T) \Omega_{\hat T}^2}{1 + D(\hat T) \Omega_{\hat T}}
\end{equation}  
Secondly, the coefficients can be determined by equating the Pad\'e expansion in the small $\Omega_{\hat T}$ and 
large $\Omega_{\hat T}$ limits with the corresponding Feynman-Kac expectations of simple polynomials.  
The details of this calculation are in Appendix A, and yields  
values of the coefficients $A,B,C$ and $D$ as given in Eq. (\ref{eq:padecoeffs}) below.
Again note that even in this general case, the naive approximation of Eq. (\ref{eq:approx}) yields a 
slightly different result, yet numerically the two are practically indistinguishable. 
This can again be understood by the plots shown in Appendix A, comparing the two approximations with Monte-Carlo 
simulations. Note also that the Pad\'e approximation of order 2:1 which we use in Eq. (\ref{eq:xhatT}) is already  
extremely close to the Monte-Carlo results (see Appendix A, Figure 12b); yet if desired, convergence can easily 
be further improved simply by including higher order terms.

\section{European Call}

The final expression for the stock price at time $T$ 
with respect to the martingale noise is thus given by 
Eq. (\ref{eq:ST}) together with Eq. (\ref{eq:xhatTPade}).
With this result, it is straightforward to obtain a general 
expression for the price of a European call 
in the current framework.
The price of  a European claim $c$ can formally be written as
\begin{equation}
c = e^{-rT} \langle h(S_T) \rangle_Q
\end{equation}
where $h$ is the payoff of the option. For a European call the payoff is $h = \max(S_T-K, 0)$, 
allowing us to evaluate the fair call  price as
\begin{equation}
\label{eq:cp}
c=  e^{-rT} \langle S_T - K) \rangle_Q^D
\end{equation}
where $D$ represents the domain of non-zero payoff, namely 
\begin{equation}
\label{eq:callcondition}
S_T > K
\end{equation}
In our current framework there must be an additional constraint on the paths expiring in the money, namely that they never crossed
$S(t) = 0$ at any $t< T$. The probability $\tilde{p} = P(S(T)>K \mid S(t) \le 0)$ that any path would cross $0$ and then return to expire greater than $K$ is intuitively  the order of the probability of default squared. More exactly though,  
it can be computed in the limit where $\eta=\sigma (1-\alpha)$ is
small, corresponding to small default probabilities ${\cal P}$. Using most probable path 
methods \cite{usunpub}, we find 
\begin{equation}
\label{eq:tildep}
\tilde{p} \approx {\cal P} \times \exp \left(-\frac{b}{T\eta^2}\right),
\end{equation}
 where $b$ 
is a positive constant of order unity. For practical purposes, 
this difference is extremely small when compared to the accuracy of our
solution, and can be neglected in most cases.

The  condition $S_T=K$ implies a  quadratic equation 
for $\Omega_T$ which can easily be solved, 
resulting in two roots $d_1$ and $d_2$ given in Appendix A, Eq. (\ref{eq:roots}). These define
a domain of integration for which the inequality $S_T>K$ is satisfied.
It can be seen  
that $d_1$  always corresponds to the relevant lower root, while the upper bound is
$d_{max}=d_2$ if $d_2 > d_1$, and $d_{max}=\infty$ otherwise. (Note that the strictness of this upper bound is however in principle  irrelevant because of the smallness of the probability distribution in that region, and
also because our approximation scheme breaks down in that region). 

We are now ready to state one of our main results, namely the closed-form expression for the price of 
a European call option within this framework. 
Following Eq. (\ref{eq:cp}), it is given as
\begin{eqnarray}
\label{eq:callskew}
c &=& S_0  \int_{d_1}^{d_{max}} (1 + (1-\alpha)x(\hat{T}))^{\frac{1}{1-\alpha}} P_q(\Omega_{\hat{T}}) d\Omega_{\hat{T}} 
\nonumber\\& & \nonumber\\
&-& e^{-rT}K\int_{d_1} ^{d_{max}} P_q(\Omega_{\hat{T}}) d\Omega_{\hat{T}}
\end{eqnarray}	                 
with $x_{\hat{T}}$ a function of $\Omega_{\hat{T}}$ given by Eq. (\ref{eq:xhatTPade}) evaluated at $u=\hat{T}$, $\hat T$ is given
by Eq; (\ref{eq:that}) with $t=T$, and
$P_q(\Omega_{\hat{T}})$ is the probability of the final value of $\Omega$, given by:
\begin{equation}
\label{eq:pq}
P_q = \frac{1}{Z(\hat{T})}(1+(q-1)\beta(\hat{T})
\Omega_{\hat{T}}^2)^{-\frac{1}{q-1}}.  
\end{equation}
Note that it is at this point that we are in fact 
overcounting those paths which crossed zero yet expired above $K$. 
To account for this, $P_q$ should 
be replaced with $P_q - \tilde{p}$ of Eq. (\ref{eq:tildep}). However, 
as mentioned above, 
$\tilde{p}$ can readily be dropped as it leads to negligibly 
small contributions, and the above result for the option price 
can be seen as exact to order $\sigma^2$. 

\section{Special Cases}

As already mentioned, several special cases are recovered for certain values of $q$ and $\alpha$. 
In practice, occasions could arise in which it  might be useful to work with these simpler solutions. 
Therefore, we spend a few lines discussing them here.

\subsection{CEV model q=1} 

The case of $q=1$ and general $\alpha$ corresponds to the CEV model of Cox and Ross \cite{Cox&Ross}. One obtains 
for the call price
\begin{equation}
\label{eq:calphaq1}
c =  S_0 \int_{d_1}^\infty (1 + (1-\alpha) x(\hat{T}))^{\frac{1}{1-\alpha}} 
P(z_{\hat{T}})dz_{\hat{T}} - e^{-rT}K\int_{d_1}^\infty P(z_{\hat{T}}) dz_{\hat{T}}
\end{equation}
with notation  $z_t = \Omega_{q=1}(t)$ and where
\begin{equation}
x(\hat{T}) = - \frac{\alpha}{2} \sigma^2{\hat{T}} + \Gamma
\end{equation}
with
\begin{equation}
\Gamma = 1 + \frac{\sigma^2\alpha(1-\alpha)}{2} \gamma \hat{T}
\end{equation}
and
\begin{equation}
 d_1 = \frac{1}{\sigma \Gamma} \left(\frac{ (Ke^{-rT}/S_0)^{1-\alpha} - 1}{1-\alpha} + 
 \frac{\alpha \sigma^2 \hat{T} }{2}\right)
\end{equation}
Of course, $z_t$ is a Gaussian variable so
\begin{equation}
P(z_t) = \frac{1}{\sqrt{2 \pi t}}\exp{-\frac{z_t^2}{2t}}
\end{equation}
These equations look different to the solutions presented in \cite{Cox&Ross}, which are expressed in terms of 
Bessel functions. Numerically, however, our solution and theirs are the same. 
The reason that they look different 
is -- apart from the small $\sigma^2$ approximation we have made -- is that 
we have made a point of explicitly averaging with respect to the distribution 
of the random noise variable $z$ 
rather than with respect to the distribution of the stock price itself. 
The reason for this is because as we saw above in the general case, 
we know something about the distribution of 
the noise $\Omega$, whereas the distribution of $S$ itself 
is more complicated -- but see Appendix B.  

\subsection{Non-Gaussian Additive or q-Normal Model $\alpha = 0$}
 
The CEV model with $\alpha=0$ is known as the {\it{normal model}}. Here we present the solution of 
the corresponding model for general $q$, which we term the q-normal model. In this particular case, 
it is easy to solve for the option price in terms of the distribution of $S$, since the noise is additive.
The call price reads:
\begin{eqnarray}
c & = &\frac{e^{-rT}}{Z(T)\sigma}\int_K^{\infty} S \, \left[1+(q-1)\frac{\beta(T)}{\sigma^2}(S-S_0e^{rT})^2
\right]^{-\frac{1}{q-1}}\, dS
\nonumber \\
&-&
 \frac{K e^{-rT}}{Z(T) \sigma}\int_K^{\infty} \left[1+(q-1)\frac{\beta(T)}{\sigma^2}(S-S_0e^{rT})^2
\right]^{-\frac{1}{q-1}}\, dS
\end{eqnarray}

\subsection{Non-Gaussian Multiplicative $\alpha = 1$}

Finally, the case studied in \cite{qf_borland} is recovered for general $q$, and $\alpha = 1$. For completeness 
we cite this result, albeit using the more precise Feynman-Kac evaluation of the path integral which resulted in Eq. 
(\ref{eq:stoldfk}):
\begin{eqnarray}
c
&  = &  S_0\int_{d_1}^{\infty} \exp\left\{\sigma \Omega_T -\frac{\sigma^2}{2} [\gamma(T)
\frac{3-q}{2}- (1-q)(g_0(T) + 
g_2(T)\Omega_T^2)] \right\} \, \nonumber \\
& & P_q(\Omega_T)  d\Omega_T 
  -  e^{-rT} K \int_{d_1}^{\infty} P_q(\Omega_T) d\Omega_T 
\end{eqnarray}
with $P_q$ as in Eq. (\ref{eq:pq}) and $d_{1}$ is the smallest root of the equation Eq. (\ref{eq:stoldfk}) such that
$S(\Omega_T)=K$.

\section{Numerical Results}

Now that we have pricing formulas for a non-Gaussian model with skew, it is interesting to look at 
the option prices which result from the model, and to see how they compare with a standard Black-Scholes
formalism.  Using Eq. (\ref{eq:callskew}) with  $q=1.5$, $\alpha = -1.5$, $\sigma = 30 \%$, $r = 6 \% $ and $S_0 = \$ 50$, 
we calculated call option prices as a function of different strikes $K$ and times to expiration $T$. 
We then backed out ìimplied volatilities, i.e. the value of $\sigma$ which would have been needed in conjunction with 
the standard Black-Scholes 
model ($q=\alpha=1$) to reproduce the same option prices. A plot of those implied volatilities as a function of 
$K$ and $T$ constitutes the skew surface which we show in Figure 4. The general shape of this surface is very similar 
to what is observed on the marketplace. For small $T$, the profile across strikes is smile-like, 
becoming more and more of a downward sloping smirk as $T$ increases.  In Figure 5 we show a similar plot 
for $q=1$ (i.e. the CEV model). Note that this skew surface does not
capture the shape one observes empirically. This goes to show that both fat tails ($q >1$) and skew ($\alpha < 1$) are
necessary for a realistic description.

\begin{figure}[t]
\psfig{file=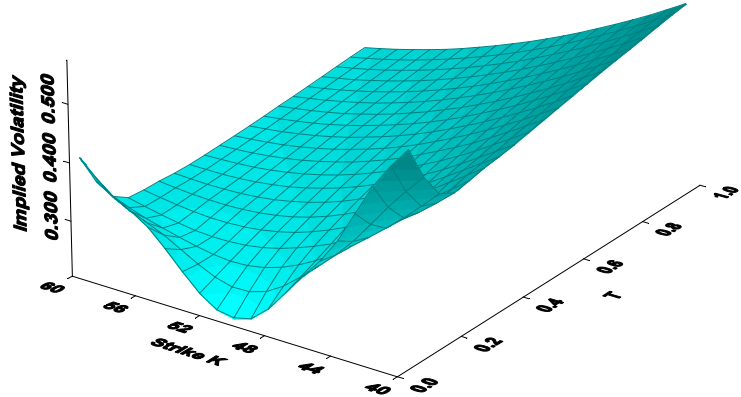,width= 4.5 in}
\caption{\footnotesize
A plot of the skew surface, i.e. Black-Scholes implied volatilities across strikes ($K$) and 
time to expiration ($T$), backed out from our non-Gaussian model with $q = 1.5$ and $\alpha = -1.5$. 
Other parameters were $S_0 = \$50, r=6 \% $ and $\sigma =30 \%$.
}
\end{figure}

\begin{figure}[t]
\psfig{file=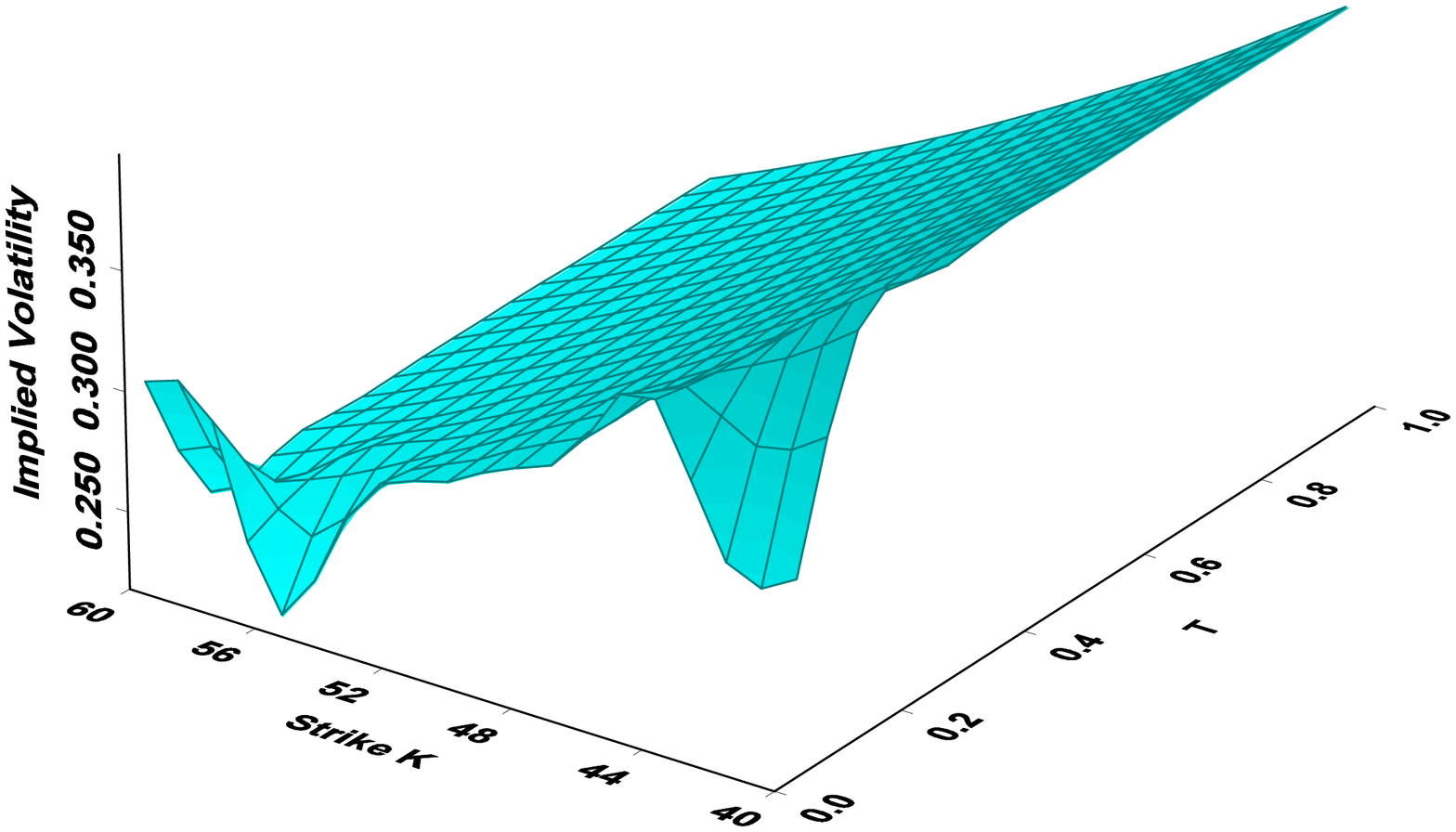,width= 4.5 in}
\caption{\footnotesize
 The skew surface implied from a CEV model  ($q = 1$) with $\alpha = -1.5$. 
Other parameters were $S_0 =\$ 50, r= 6 \%$ and $\sigma =30 \%$. The kink at short times may just be a numerical artefact.
}
\end{figure}

It is also interesting to look at the variation of the option price with respect to the new parameter $\alpha$
(the variations to q, denoted by Upsilon $\Upsilon$, were studied in \cite{qf_borland}). We denote this new `Greek' 
by the Hebrew letter Aleph $\aleph$, such that $\aleph = \partial c/\partial \alpha$. Figure 6
shows $\aleph$ as  a function of stock price $S_0$, using $K= \$ 50$, $T=0.5$ years, 
$\sigma = 30 \%$, and $ r = 6 \%$, for $q=1.5$ and $q=1$.
The asymmetric nature of the sensitivity to variations in Aleph is apparent for $q=1.5$. 
Another way of depicting much the same information is shown in Figure 7, where volatilities
implied by comparing standard Black-Scholes with our model Eq. (\ref{eq:callskew}) for different values of $\alpha$ 
are shown. In all cases we used $q=1.5$, $S_0 = 50$, $T=0.5$ years, $\sigma=30\%$ and $r=6\%$. One can clearly see that the
implied volatility curve is like a smile for $\alpha = 1$ (no skew) becoming more and more asymmetric about $K=S_0$ as 
$\alpha$ decreases.

\begin{figure}[t]
\psfig{file=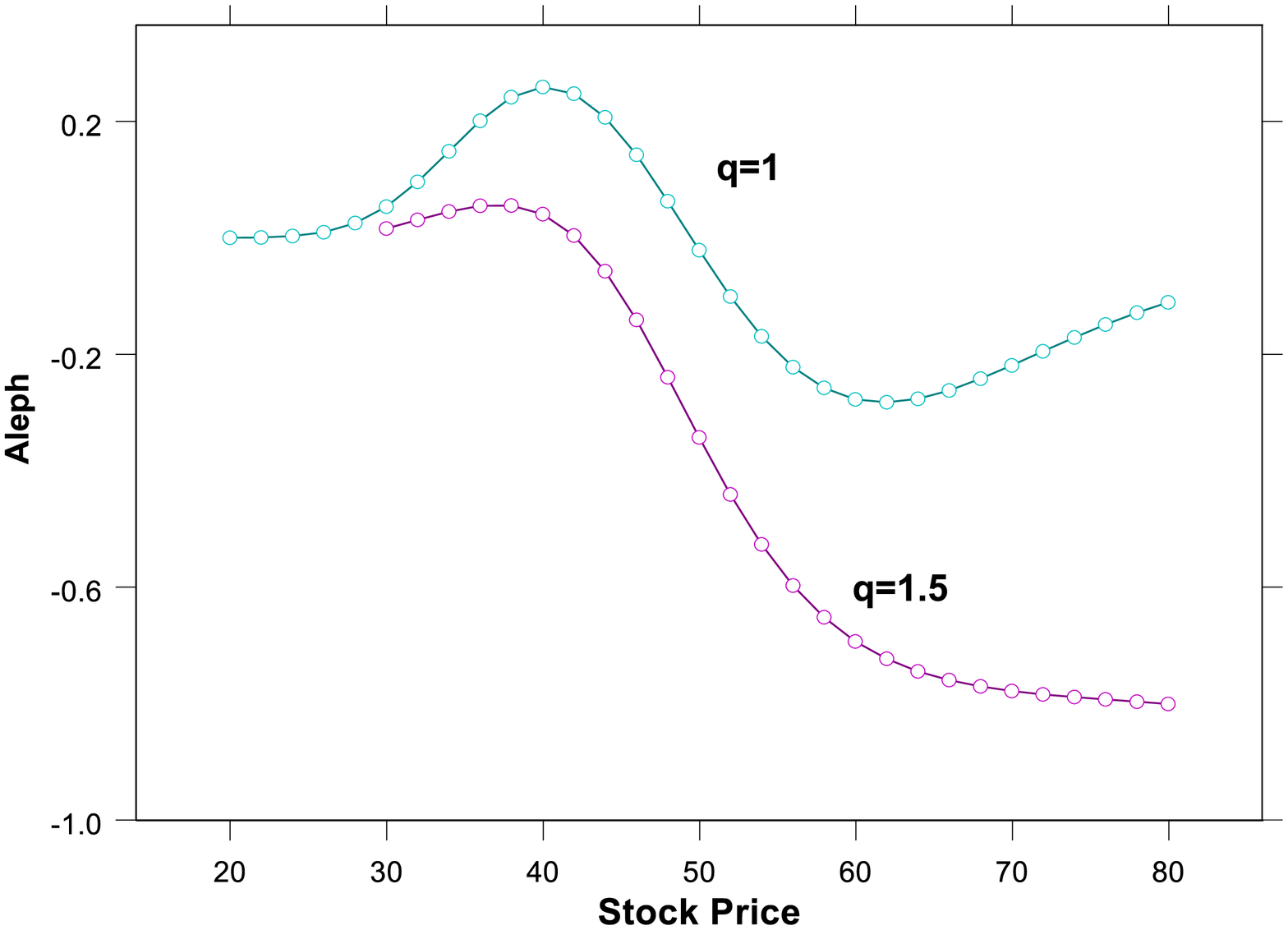,width= 4.5 in}
\caption{\footnotesize
Aleph the Greek, $\aleph$: The partial derivative of the call price with respect to 
$\alpha$ is plotted as a function of
the stock price $S(0)$, for $q=1.5$ and $q=1.$. The asymmetric nature is apparent for $q=1.5$. 
We used $K =\$50, r=6\%,\sigma =30\%$ and $T=0.5$ years.
 }
\end{figure}

\begin{figure}[t]
\psfig{file=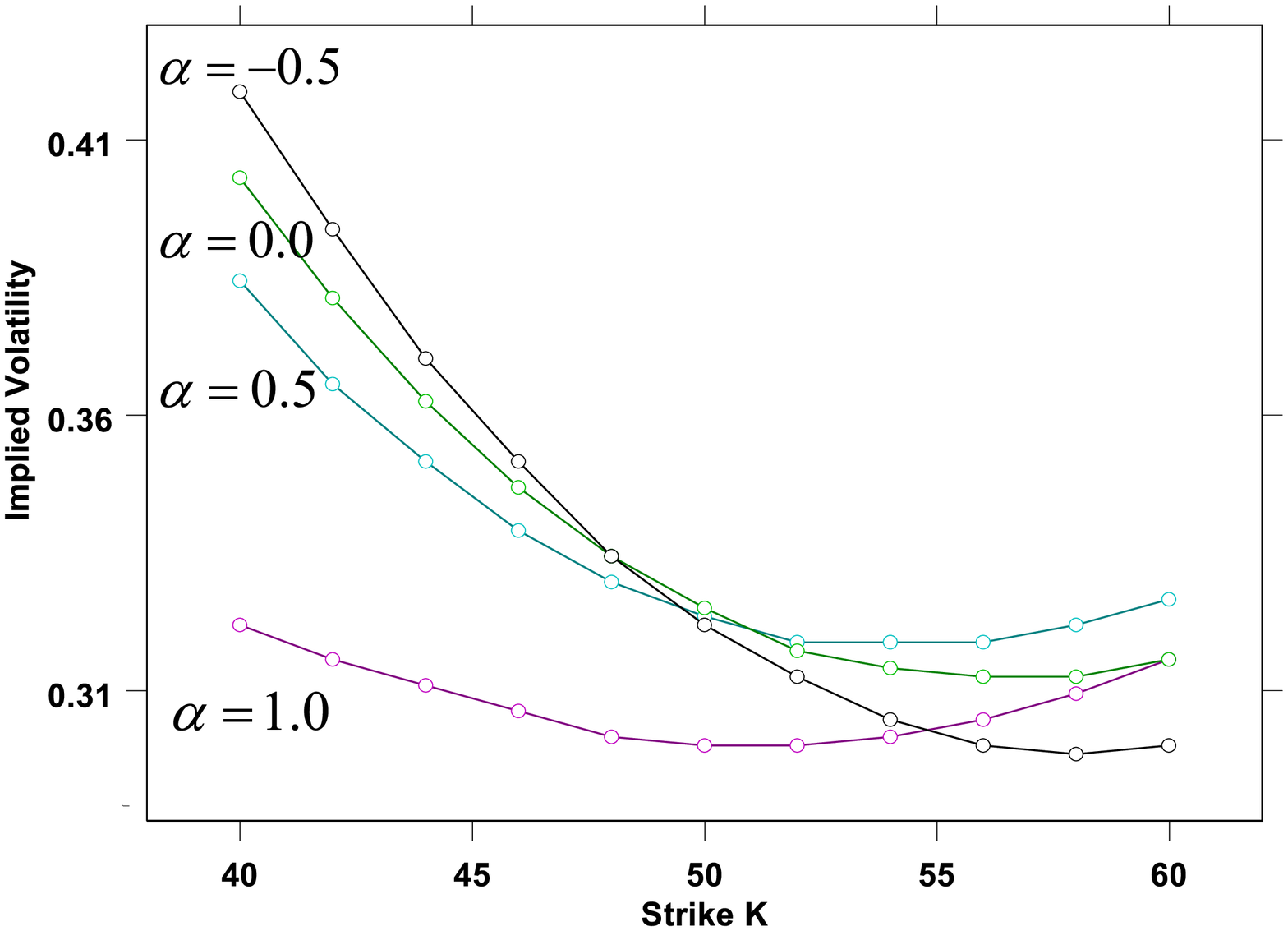,width= 4.5 in}
\caption{\footnotesize
 Implied volatilities from the $q=1.5$ model as a function of $\alpha$,
with $S_0 =\$ 50 , r=6\%,\sigma =30\%$, and $T=0.5$ years. As $\alpha$ decreases,  the skew increases.
Intuitively this make sense: $\alpha$ contributes to larger negative tails in the distribution of the underlying, 
and therefore there will be higher probability of expiring out of the money 
relative to a Black-Scholes process with lognormal noise. For extreme out of the money values 
however, the noise in the fat tails can again increase the  probability that an out-of-the money 
option can expire in the money, so we expect the smile to increase again in this regime.    
 }
\end{figure}

In all of these results, we used the closed form pricing formula, Eq. (\ref{eq:callskew}), to generate option prices.
However, since we used some approximations along the way, 
it is a good check to see how the closed form
price compares with that obtained from pricing via Monte-Carlo simulations of the process. Indeed, we saw that 
the two values are indistinguishable within the limits of accuracy of the Monte-Carlo simulations.

%\begin{figure}[t]
%\psfig{file=Fig7.ps,width= 4.5 in}
%\caption{\footnotesize
% Option prices obtained via Monte-Carlo simulations are compared with those of the closed-form solution. Values
%used were $q=1.5$, $\alpha=-1$, $\sigma=30\%$, $r=6\%$ and $T=0.5$ years.
% }
%\end{figure}

\section{Empirical Results}

To illustrate the qualitative agreement between our model and market data, we show in 
Figure 8 a plot of 
the empirical skew surface for OEX options, as well as the skew surface backed out of our model with $q=1.5$ 
and $\alpha = -1.2$. We have not at all tried to calibrate the model to match the empirical data in any way, 
the plot is only intending to show that 
our model produces a surface with similar features to the empirical one across several time scales, 
with just one set of parameters $\alpha,\sigma$ and $q$. In other words while with respect to the Black-Scholes 
model, the entire skew surface
of Figure 8b  is needed to describe the option prices, with respect to the $q=1.5$ and $\alpha=-1.2$  model the 
skew surface reduces to a single point. Consequently, we have the hope that real market smiles and skews 
might be captured by our model
with parameters $q,\alpha$ and $\sigma$ varying only slightly across strikes and times 
to expiration.

\begin{figure}[t]
\psfig{file=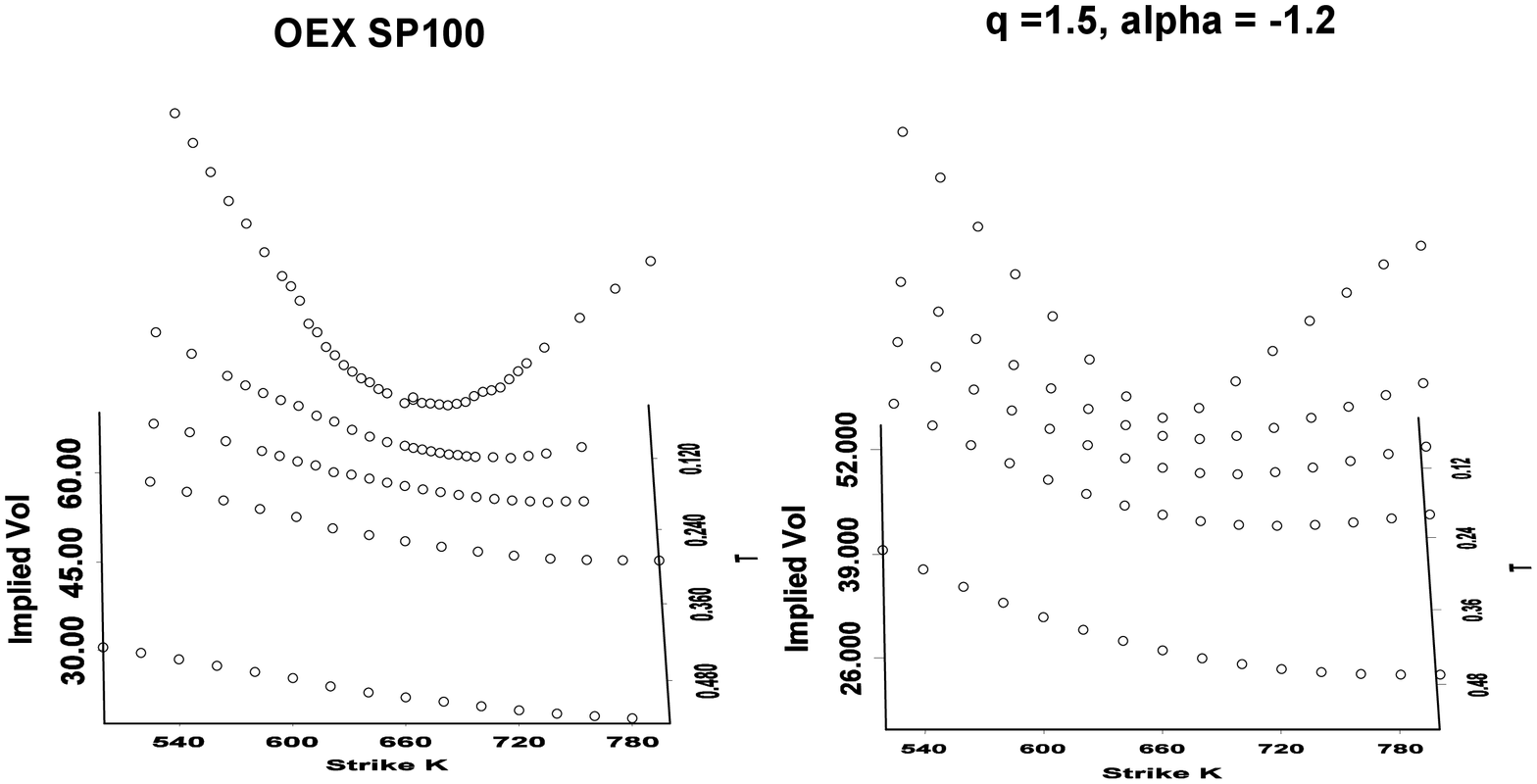,angle=270,height= 4 in}
\caption{\footnotesize
  A purely qualitative comparison between the empirical skew surface of a set of OEX options on S\&P100 futures
traded on June 6, 2001, and the implied skew surface from our model with $q=1.5$, $\alpha=-1.2$, $\sigma= 30\%$,
$r = 4.5 \%$ and $S_0 = 660$. We have not tried to 
calibrate to the OEX data in any way, we simply wish to show that the general behaviour of the surfaces across 
strikes and time to expiration is similar. From top to bottom: $T=0.03$, $T=0.12$, $T=0.20$, $T=0.29$ and $T=0.55$.
 }
\end{figure}

Clearly, to test this 
with any statistical significance requires a large study on many options, which we leave for a future work. 
However, we do present, again for the purpose of illustration, results based on an analysis 
of one set of options, namely call options on Microsoft (MSFT)
traded on November 19, 2003. A popular methodology  that market makers and traders follow is to  
vary  the parameters of whatever model they are using  such that the
theoretical smile matches the market at each time to expiration $T$.  
 We could also follow such an approach: vary $q$, $\alpha$ and $\sigma$ for 
each value of $T$ such that the smiles and skews  are reproduced.  
But if the model is good in the sense that the parameters do not change much over time, 
then perhaps the most parsimonious treatment would be  to choose one set of those parameters such that the 
entire skew surface across both strikes and expiration times 
is well-fit. While interesting, both of these approaches would be ways of {\it implying} 
the model parameters from the options data. 

An alternative  methodology which is well-suited for our current approach, is to instead  try to relate at 
least one or two of the model parameters to properties of the underlying asset, and then use these to 
{\it predict} market smiles.  In particular, the parameter $q$ can be readily determined from the empirical distribution 
of the returns of the underlying asset. It has been found in previous studies \cite{osorioetal,qf_borland,Michael&Johnson} 
that 
a value of $q \approx 1.4$  captures well the distribution of daily stock returns, so this value could be adopted 
in the option pricing formula. The parameter $\alpha$ could in principle also be determined from the distribution 
of underlying returns, or from measuring a leverage correlation function \cite{leverage,jpbouchaud}. Alternatively, 
it could 
perhaps be determined  (as mentioned earlier in this paper) form empirical probabilities of default. For the present 
study we fix only $q$ from the underlying distribution, and imply $\alpha$ and $\sigma$ from the market smiles.
 We shall then look at how good the implied smiles fit the data, 
in conjunction with how the implied parameters vary with $T$. If they are somewhat stable then we can conclude 
that the model is quite good.  

In Figure 9 we show the results for our model. As $T$ ranges from the order of a month to a year, we choose 
$\alpha$ and $\sigma$ such that
the best fit in terms of least square pricing error is obtained between model and empirical call prices, for fixed $q$.
The plots show actual implied volatility
and the implied volatility of our model. 
It is clear to see that the $q=1.4$ model provides a good fit of observed smiles at each time to expiration. However
for the longest maturity $T=1.17$ one sees that the away-from-the money strikes 
are slightly over-valued by our model. 
The valid question remains as to whether this discrepancy is a true market mispricing 
or an artifact of the model, which does not lead to log-normal statistics for large $T$. 
(Note that we can easily match the market exactly even at $T=1.17$ 
simply by reducing $q$ appropriately,
but this is not what we are attempting to do here.)

\begin{figure}[t]
\psfig{file=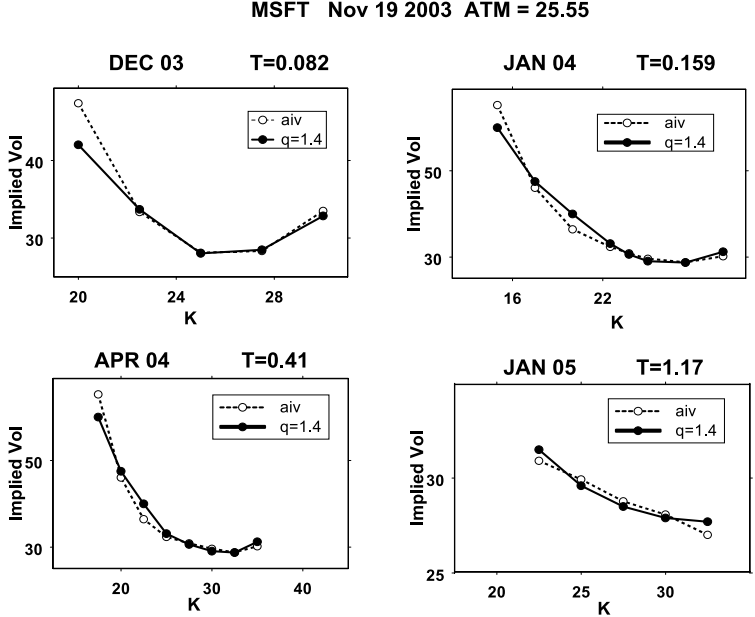,width= 4.5 in}
\caption{\footnotesize
 A quantitative comparison between the empirical skews obtained from a set of MSFT 
options traded on 
November 19, 2003, and our model with $q=1.4$, which well-fits the returns distribution of the underlying stock.
  We varied $\alpha$ and the volatility parameter $\sigma$ of the $q=1.4$ model for each time to expiration $T$ 
  (see Table 1). We used $r = 4.5 \%$ and $S_0 = \$25.55$. AIV stands for the actual average (of put and call) implied volatility.
 }
\end{figure}

\begin{table}[h]
\begin{center}{
\begin{tabular}{|l|c|c|c|c|}
\hline
T [Years] & 0.082 & 0.159 & 0.41 & 1.17\\  \hline
$\sigma$ [\%]& 32 & 31 & 27 & 25 \\ \hline
$\alpha$ & 0.1 & 0.2 & 0.3 & 0.2\\
\hline
\end{tabular} }
\end{center}
\caption{\footnotesize Variation of the fitted parameters $\sigma$ and $\alpha$ with time to expiration $T$, for $q=1.4$.}
\label{turns}
\end{table}

In Table 1 we show how the parameters of the  model vary as a function of $T$.
The parameter $\alpha$  goes from $0.1$ at early times to fluctuate around $0.2$. 
Note that this order of magnitude of $\alpha$, when related back to the distribution of stock returns, 
is  entirely consistent with empirical observations  of the leverage effect \cite{leverage}. 
The volatility parameter $\sigma$ exhibits a negative term structure, 
ranging from $32\%$ to $ 25\%$.  
Such a negative term structure has been consistently observed in our empirical studies 
(for example it is seen in ref. \cite{qf_borland} where options on FX futures are analyzed). 
The reason for such an effect is, as mentioned earlier,  that the 
$q$-model considered here predicts an anomalous growth of 
the volatility with time, as $T^{1/(3-q)}$, instead of the empirically observed standard diffusive scaling, 
$\sqrt{T}$. A way to correct for this is to allow the volatility $\sigma$ to be maturity dependent, which 
amounts to a mere redefinition of time in the nonlinear Fokker-Planck equation that defines the model. 
For $q=1.4$, $\sigma$ should scale as $T^{-1/8}$ to reproduce the correct time dependence of the 
width of the return distribution with time. This scaling appears indeed to be satisfied as shown in Figure 10, where we 
plot $\sigma$ (Table 1) of the MSFT data against time on a log-log scale. 
To illustrate the regularity of this temporal behaviour, we also show in Figure 10 the corresponding plot for the 
$\sigma$ parameter of the options on FX futures initially shown in \cite{qf_borland}.
Based on these results,  we see that the total variation of the model parameters $\alpha$ and the maturity rescaled 
$\sigma$ are only very slight. Furthermore, by construction $q$ is kept constant.

Another comment, which should be taken loosely since we have not done a systematic study 
of this point, concerns the probabilities of default implied by  the parameters of this 
MSFT example, when reinserted  in our stock price  model. Assuming that the real drift of 
MSFT is the risk free rate (which is certainly an underestimate), one obtains 
 $< 0.005 \%$ probability of bankruptcy  for $T=.082$, $0.01 \%$ for 
$T=0.159$, $0.07 \%$ for $T= 0.41$ and $0.35\% $ for $T=1.17$. 
These estimates were obtained from Monte Carlo simulations, and do not seem unreasonable. 
However, the above probability of default at 1 year is probably higher that 
typical credit risk ratings of MSFT. This overestimation of the probability of default 
at longer timescales is intimately related to the smile shown in Figure 9, for $T= 1.17$: 
the away-from-the money strikes are slightly over-valued by our model as $T$ increases. 
As mentioned earlier, this is because we choose to keep $q=1.4$ fixed at all timescales. If we 
instead decided to vary $q$ to fit the smile exactly, it would be closer to $q=1$ 
and the default probabilities would drop much closer to $0$ again (for example, see Figure 3 b).
Also note that  in reality, default probabilities might depend on the real rate of return of 
the stock, which for MSFT is substantially higher than the risk free rate.

\begin{figure}[t]
\psfig{file=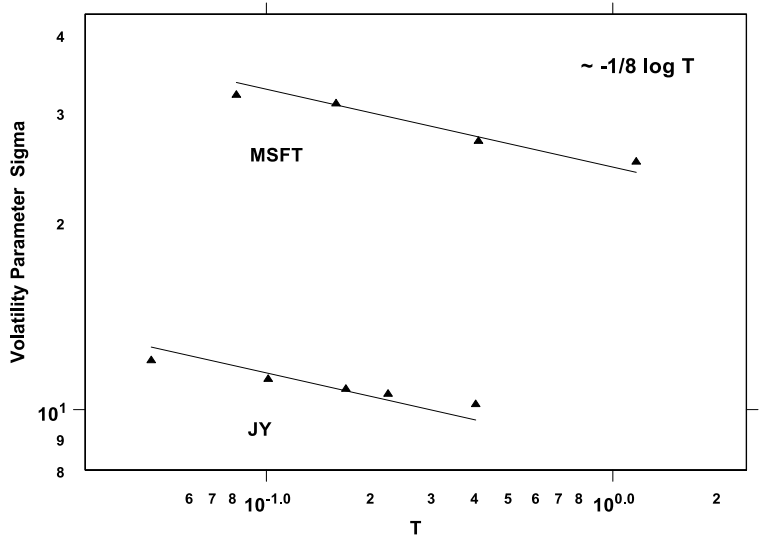,width= 4.5 in}
\caption{\footnotesize
Deterministic term-structure in the volatility parameter $\sigma$ of the $q=1.4$ model, shown (top) for the
MSFT options example of Figure 9 and Table 1; shown (bottom) for the JY Futures options example discussed in \cite{qf_borland}.
   }
\end{figure}

It could be interesting to  place our results in the context of a comparison with a popular stochastic volatility model 
(namely the SABR model, which for the sake of  this discussion is briefly summarized in Appendix C).  
The fit of the SABR model to the MSFT options discussed above was performed by an external source 
\cite{source} so we only briefly report the results here.
Three parameters were varied at each $T$, and consequently the smiles could be fit much as in our example. 
The volatility parameter $\sigma$ varied with a slightly positive term structure around
a value of  $29\%$. However the parameters $\rho$ (related to the correlations between stock fluctuations and
 volatility correlations) and $\lambda$ (which describes the volatility of the volatility) did vary quite  a bit.
We saw that $\rho$
increased from $0.1$ to $0.6$. Perhaps more significantly, $\lambda$ went from around  $9.15$ to $1.14$ as maturity increased
from $0.082$ to $1.17$. This decrease makes sense because at large times, 
the stochastic volatility in the SABR model is unbounded; consequently the implied option smile would be way 
too pronounced if  $\lambda$ did not suppress this feature, forcing the model 
into the log-normal limit as $T$ increases. On the whole, we saw that while the SABR model certainly 
fit the market smiles well, the number of parameters (3 if $\beta$ is fixed) and their variation is larger 
than in the non-Gaussian $q=1.4$ model (with 2 free parameters).

Consequently, we can conclude 
(at least for the options studied in this example), that our non-Gaussian model with fixed $q$ indeed yields 
a relatively parsimonious  description of empirically observed option prices, capturing both the smile and the skew with a 
few seemingly robust parameters.
  
\section{Conclusions}

In this paper we have extended the non-Gaussian option pricing theory of \cite{qf_borland} 
(which recovers the standard Black-Scholes case in the limit $q=1$) to include asymmetries in the underlying process. 
These were introduced so as to incorporate the leverage correlation effect \cite{jpbouchaud} in a fashion such
that the CEV model of Cox and Ross \cite{Cox&Ross} is recovered in appropriate limits ($q=\alpha=1$). 
A theoretical treatment of the problem is possible
much along standard lines of mathematical finance. Using the fact that the volatility of the process is a
deterministic function of the stock value, the zero-risk property of the Black-Scholes hold, and one can 
set up a generalized Black-Scholes 
PDE as well as define a unique martingale measure allowing us to evaluate option prices via risk-neutral expectations. 
We have introduced a generalized Feynman-Kac formula for the family of stochastic processes involved in our model, namely
statistical feedback equations of the type \cite{pre_borland} which evolve according to a nonlinear Fokker-Planck equation
\cite{Tsallis&Bukman},\cite{Chasseigne&Vasquez}. This formula in conjunction with Pad\'e expansions could then be used to 
evaluate closed-form option pricing equations for European call options. 

Numerical results allow us to back out Black-Scholes implied volatilities. A plot of those across strikes $K$ and 
time to expiration $T$ constitute a skew surface. We found that the skew surfaces resulting from our model exhibit 
many properties seen in real markets. In particular the shape of the surface tends to go from a pronounced smile 
(across strikes) to a sloping line as $T$ increases.  A comparison of the model to a restricted set of MSFT options 
was discussed. We found that the entire skew surface could be well-explained with a fixed value of $q=1.4$, which 
also fits to the distribution of the underlying. The remaining parameters $\alpha$ and $\sigma$ vary only slightly, 
especially after we factored out a deterministic term structure in the volatility parameter. 
While the philosophy of many market participants is in general to tune the parameters of their models to fit market smiles, 
we try to relate our parameter to the underlying distribution in the hope that we can attain a more parsimonious 
(and therefore more stable) description of both underlying and option. (Along this line of thought, see also 
\cite{hedged} and refs. therein). Indeed, our results -- though not yet 
statistically significant -- do indicate that it  might be possible to explain the entire skew surface with one 
set of constant (or slowly varying) parameters. 
We hope to strengthen this statement through  more  empirical studies.

\bigskip
{\bf Acknowledgements: } We wish to thank Jeremy Evnine, Roberto Osorio and 
Peter Carr for interesting and useful comments.  Benoit Pochart is acknowledged for 
careful reading of the manuscript. 

\bigskip

\section{Appendix A: Feynman-Kac Expectations and Pad\'e Coefficients}

The coefficients of the Feynman-Kac Ansatz Eq. (\ref{eq:fkAnsatz}) when inserted into Eq. (\ref{eq:FeynmanKac}) must satisfy
\begin{eqnarray}
\label{eq:fkcoeffsf}
\frac{dg_0}{du} &=& Z(u)^{1-q}g_2(u) \\ \label{eq:fkcoeffsn}
\frac{dg_1}{du} &=& 2(q-2)Z(u)^{1-q}\beta(u) g_1 + h_1(u)\\ \label{eq:fkcoeffsg}
\frac{dg_2}{du} &=& (5q-9) Z(u)^{1-q}\beta(u) g_2(u) + h_2(u)
\end{eqnarray} 
From these equations follows  that for path integrals of type 
$\exp\{ \int_0^u h_1(t) \Omega(t) dt \}$ (with $h_2=0$), then only the $g_1$ coefficient is relevant 
and the expected value of the integral yields the approximation
\begin{equation}
\int_0^u h_1(t) \Omega(t) dt = g_1(u) \Omega_u
\end{equation}
Similarly, if we are only looking at path integrals of the type
$\exp\{ \int_0^u h_2(t) \Omega(t)^2 dt \} $ (with $h_1=0$), coefficients $g_0$ and $g_2$ are coupled together, 
so that the Feynman-Kac approximation of the integral implies
\begin{equation}
\label{eq:fkomega2}
\int_0^u h_2(t) \Omega(t)^2 =  g_0(u) + g_2(u) \Omega_u^2 
\end{equation}

In the option pricing problem without skew, namely the case $\alpha=1$ of Eq. (\ref{eq:STalpha1}), 
we have $h_1 =   0$ and $h_2 = Z^{1-q}\beta$. Integration yields
\begin{eqnarray}
g_0(u) &= &\gamma(u) \frac{3-q}{2(9-5q)} \label{eq:fkf}\\
g_2(u) & = & \frac{1}{9-5q} \label{eq:fkg}
\end{eqnarray}
with
\begin{equation}
\label{eq:gammaT}
\gamma(u) =((3-q)(2-q)c_q)^{\frac{q-1}{3-q}}u^{\frac{2}{3-q}}
\end{equation}

For the general case of option pricing including skew (general $\alpha$ as in Eq. (\ref{eq:xhatT})), we expand the 
Pad\'e Ansatz of Eq. (\ref{eq:xhatTPade}) for both small and large  $\Omega$, and equate
with the same expansions of the actual quantity we are interested in, which we then evaluate
using Feynman-Kac expectations. In the small $\Omega$ case:
\begin{eqnarray}
A &+& (B-AD)\Omega_u + (C-BD +AD^2)\Omega_u^2 \\ &=&
\int_0^{{u}} Z(t)^{1-q}\left[1-\eta \Omega(t) + (1-q) \beta(t) \Omega(t)^2 \right] dt  \nonumber \\
& =&  \int_0^{{u}} \left[ Z(t)^{1-q}  -\eta h_1(t) \Omega(t) + (1-q) h_2(t) \Omega(t)^2 \right] dt \nonumber\\
&=& \gamma(u)\frac{(3-q)}{2}  -\eta g_1({u}) \Omega_{{u}} + (1-q) ( g_0({u}) + g_2({u}) \Omega_{{u}}^2 ) \nonumber
\end{eqnarray}
with
\begin{equation}
\eta  = \sigma (1-\alpha)
\end{equation}
and $\gamma$ as in Eq. (\ref{eq:gammaT}).

The coefficients $g_j$ are calculated from Eq. (\ref{eq:fkcoeffsf})-Eq. (\ref{eq:fkcoeffsg}) 
with $h_1 = Z^{1-q}$ and $h_2 = \beta Z^{1-q}$ and result in:
\begin{eqnarray}
g_0 &=& \gamma(u) \frac{3-q}{2(9-5q)}\\
g_1 &=& \gamma(u) \frac{(3-q)}{4}\\
g_2 &=& \frac{1}{9-5q}
\end{eqnarray}
Similarly, the large $\Omega$ expansion yields
\begin{eqnarray}
\frac{C}{D} \Omega_u& =& \frac{1-q}{\eta} \int_0^{u }\tilde{h}_1(t) \Omega(t) dt\\
&=& \frac{1-q}{\eta} \tilde{g_1}(u) \Omega_{u}
\end{eqnarray}
with $\tilde{g_1}$ calculated from Eq. (\ref{eq:fkcoeffsn}) using $\tilde{h}_1 = \beta(t)Z(t)^{1-q}$, yielding
\begin{equation}
\tilde{g_1} = \frac{1}{2(2-q)}
\end{equation}
Through standard coefficient comparison, we have enough information to solve for the Pad\'e coefficients A,B,C and D,
resulting in
\begin{eqnarray} \label{eq:padecoeffs}
A &=& g_0(q-1)+\frac{3-q}{2} \gamma\\ \nonumber
B &=&  AD -\eta\tilde{g_1}\\ \nonumber
C &=& (q-1)\frac{\tilde{g_1}}{\eta} D\\ \nonumber
D &=& \frac{g_2(q-1)}{\frac{q-1}{\eta}{\tilde{g}_1} +\eta{g}_1}
\end{eqnarray}

In the case of a general skew, the condition $S_T = K$ yields a quadratic equation with the roots
\begin{equation}
\label{eq:roots}
d_{1,2} = \frac{N \mp \sqrt{ N^2 - 4MR}}{2M}
\end{equation}
with
\begin{eqnarray}
N &=& -D\frac{ (Ke^{-rT}/S_0)^{1-\alpha}  - 1}{1-\alpha} +\sigma - B \frac{\alpha \sigma^2}{2}\\
M & =& C \frac{\alpha \sigma^2}{2} -\sigma D\\
R& =&\frac{ (Ke^{-rT}/S_0)^{1-\alpha} - 1}{1-\alpha}+ A \frac{\alpha \sigma^2}{2}
\end{eqnarray}
where $A,B,C$ and $D$ are evaluated from Eq. (\ref{eq:padecoeffs}) at the time $u=\hat{T}$ with $\hat{T}$ of
 Eq. (\ref{eq:that}).

These results are all based on evaluating the terms of type Eq. (\ref{eq:fint}) using the Feynman-Kac Ansatz
Eq. (\ref{eq:fkAnsatz}). We want to compare in details these results to the ones obtained in \cite{qf_borland} using 
a different evaluation technique, namely replacing the path $\Omega(t)$ by another path defined such 
that the ensemble distribution at each time
would be the same as for that of the original paths, namely $\Omega(t) = \sqrt{\beta(T)/\beta(t)} \Omega_T$ 
as in Eq. (\ref{eq:approx}). The numerical results obtained by that approximations are extremely close to 
those obtained by the current evaluation method, and by Monte-Carlo simulations. 
To elucidate this point we show in Figure 11a a plot of the quantity
\begin{equation}
\label{eq:test1}
\int_0^T   \beta(t)Z^{1-q} \Omega(t)^2 dt
\end{equation}
versus $\Omega_T$ evaluated for $q=1.5$ and $T= 0.5$ via i) Monte-Carlo simulations, ii) as in Eq. (\ref{eq:approx}) and iii) using the Feynman-Kac 
evaluation
Eq. (\ref{eq:oldpaperterm1}) with Eq. (\ref{eq:fkf}) and Eq. (\ref{eq:fkg}). It is clear that the approach 
Eq. (\ref{eq:approx}) slightly under-values the expectation of the true paths for small $\Omega(t)$, 
and over-values for large $\Omega(t)$, in such a way that the average value over all $\Omega(t)$ yields the correct result. 
The Feynman-Kac approximation, on the other hand, is expected to be exact in this case. 

In the more general case with a skew, we need to evaluate an expression of form
\begin{equation}
\label{eq:test2}
\int_0^T \frac{1}{1+(1-\alpha)\sigma \Omega(t)}  \beta(t)Z^{1-q} \Omega(t)^2 dt
\end{equation}
Again this can be done for $q=1.5$, $\alpha=0.5$ and $T=0.5$ using the i) Monte-Carlo simulations, ii) Eq. (\ref{eq:approx}), 
iii) Feynman-Kac equation 
together with a Pad\'e expansion. The results are shown in Figure 11b. Similar behaviour as that of Figure 11a 
is exhibited: the effective path approximation of Eq. (\ref{eq:approx}) under-values and over-values the true result 
in such a way that when averaged over all $\Omega_T$ a good approximation is obtained, whereas the Feynman-Kac approach 
is uniformly better.

The close agreement between both approximation techniques allows us in practice to use either one for option price 
evaluation. The numerical results are extremely close. Nevertheless, the Feynman-Kac approach should be preferred 
as the more exact one.

\begin{figure}[t]
\psfig{file=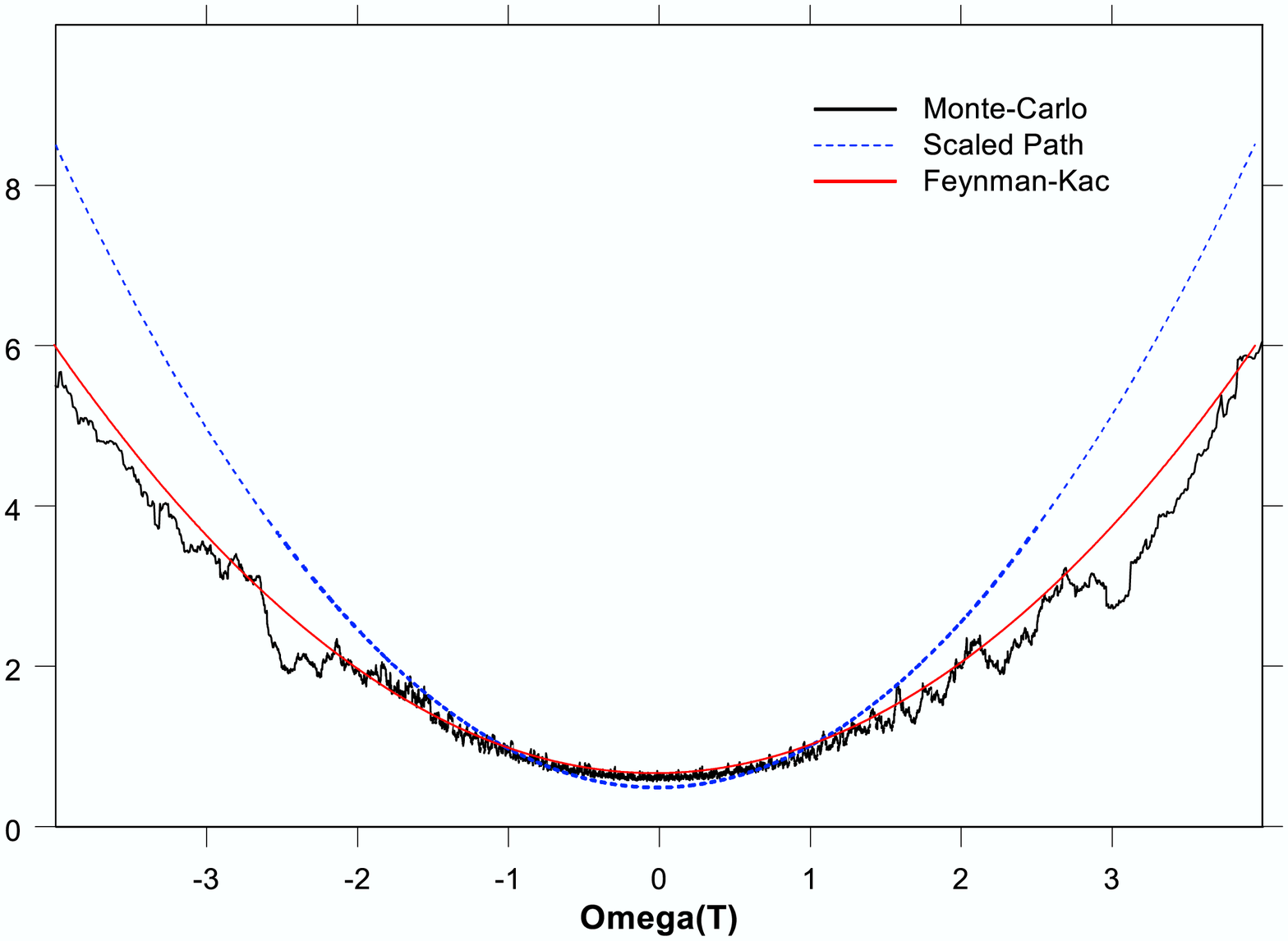,width=2.375in}\psfig{file=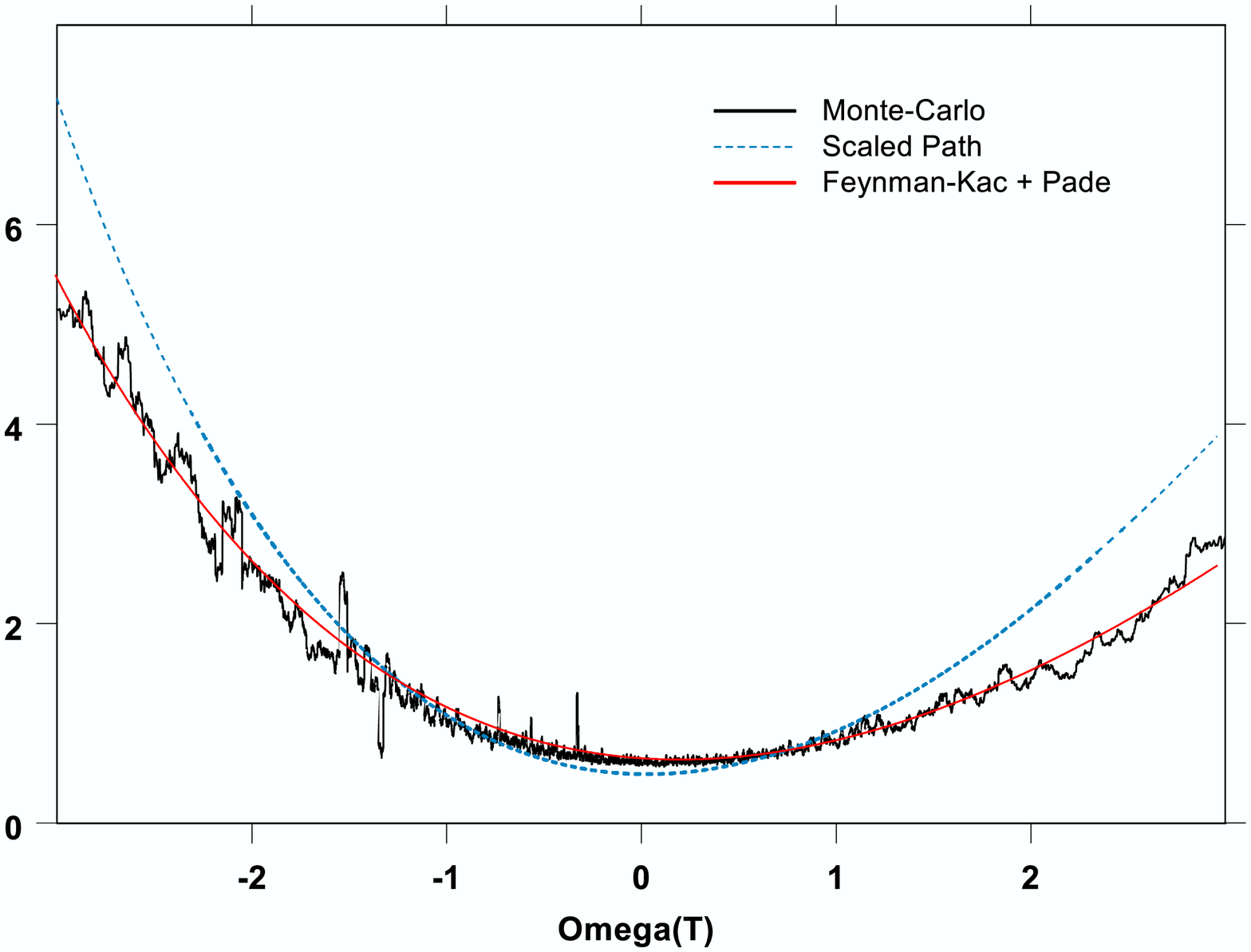,width=2.5in}
\caption{\footnotesize
Evaluation of path dependent integrals as a function of the terminal value $\Omega_T$, using i) Monte-Carlo simulations, 
ii) the approximation of Eq. (\ref{eq:approx}), and iii) Feynman-Kac techniques. 
a) The quantity Eq. (\ref{eq:test1}) relevant for $q=1.5$ and 
$\alpha=1.0$ (no skew) and b) the quantity Eq. (\ref{eq:test2}) relevant for the general skew case, 
here with $q=1.5$ and  $\alpha=0.5$. In both cases $\sigma = 30 \%$. 
 }
\end{figure}

\section{Appendix B: Exact Solutions via Hyper-Geometric Functions - A Proposal}

As an addition to  the theoretical part of our paper we would like to briefly discuss
a possible alternative path to solving for the option prices of the non-Gaussian  model with skew
 Eq. (\ref{eq:Slangevin}). The solutions will involve the fact that one can map the transformed problem of
Eq. (\ref{eq:xhat}) onto a free-particle problem in higher dimension as described below.

In the standard case $q=1$ the CEV model of Cox and Ross admits an explicit 
solution in
terms of Bessel functions for arbitrary values of $\alpha$. In this appendix, 
we show 
how this result can be obtained by mapping to CEV process to a standard 
Brownian motion
in higher dimensions, and how this method generalizes, although incompletely, to the case $q > 1$.

Starting from Eq (\ref{eq:xhat}), the change of variable $y = 1 +(1-\alpha)x$ leads 
to:
\begin{equation}
dy = \eta d \Omega - \frac{\alpha}{2(1-\alpha)}\eta^2 P^{1-q} y^{-1} dt 
\end{equation} 
with $\eta = \sigma(1-\alpha)$, in this appendix we drop the hat on the time 
variable. To order $\eta^2$, this corresponds to a Fokker-Planck equation of form 
\begin{equation} 
\frac{\partial P}{\partial t} =  a \frac{\partial}{\partial y}\frac{P^{2-q}}{y} 
+ 
\frac{\partial ^2  P^{2-q}}{\partial y^2}
\end{equation}
where the time has been rescaled by $\eta^2/2$, and $a \equiv \alpha/(1- \alpha)$.

Now, inserting the Ansatz $P = f (y,t) \Phi_0(y)$ one obtains \begin{eqnarray} \Phi_0 \frac{\partial f}{\partial t} &= &a(- \frac{1}{y^2} \Phi_0^{\nu} f^{\nu} 
+ 
\frac{1}{y}(\nu\Phi_0^{\nu-1}\frac{\partial \Phi_0}{\partial y} f^{\nu} 
+ \nu \Phi_0^{\nu}f^{\nu-1} \frac{\partial f} {\partial y})) \nonumber 
+ \\
& +&  (\nu(\nu-1)\Phi_0^{\nu-2}(\frac{\partial \Phi_0}{\partial y})^2f^{\nu} + 2 \nu^2\Phi_0^{\nu-1}f^{\nu-1}\frac{\partial \Phi_0}{\partial y}\frac{\partial 
f}{\partial y} \nonumber \\
&+& \nu \Phi_0^{\nu} (\nu-1) f^{\nu-2}(\frac{\partial f}{\partial y})^2
+ \nu \Phi_0^{\nu-1} f^{\nu} \frac{\partial ^2 \Phi_0}{\partial y^2} +
\nu \Phi_0^{\nu} f^{\nu-1} \frac{\partial^2 f}{\partial y^2}) \end{eqnarray} with $\nu \equiv 2-q$.

Grouping together all terms which do not contain derivatives of $f$, and 
imposing 
that the coefficient vanishes leads to an ordinary differential equation for 
$\Phi_0$:
\begin{equation}
\nu \frac{\partial}{\partial y}(\Phi_0^{\nu-1} \frac {\partial \Phi_0}{\partial 
y}) 
- a \frac{\Phi_0^{\nu}}{y^2} + a\frac{\nu}{y}\Phi_0^{\nu-1}\frac{\partial 
\Phi_0}{\partial y}
= 0
\end{equation}
which is solved by
\begin{equation}
\Phi_0 = y^{\lambda}
\end{equation}
provided $\lambda (\nu,\alpha)$ satisfies the following quadratic equation: \begin{equation} a\nu^2\lambda^2 + \nu \lambda (a-1) - a = 0. \end{equation} The correct root is the one that vanishes when $a \to 0$. All remaining terms 
can be regrouped as
\begin{equation}
\frac{\partial f}{\partial t} =  y^{\lambda(\nu-1)} \left( (a+2\lambda\nu) \frac {1}{y}  \frac{\partial}{\partial y}f^{\nu} + \frac{\partial ^2}{\partial y^2} f^{\nu}
\right) 
 \equiv  y^{\lambda(\nu-1)} \Delta_d f^{\nu}
\end{equation}
where
\begin{equation}
\Delta_d = \frac{\partial ^2}{\partial y^2} + \frac{d-1}{y} \frac{\partial} {\partial y}
\end{equation}
 is the radial Laplacian operator in $d$ (fictitious) dimensions, where $d$ is 
given 
by: $d-1 = a + 2\lambda \nu$.
The $y^{\lambda(\nu-1)}$ term can finally be eliminated by introducing a new 
coordinate $g = g(y)$
defined as:
\begin{equation}
g = \frac{y^{1-\lambda(\nu-1)/2}}{1-\lambda(\nu-1)/2}.
\end{equation}
In terms of this new coordinate, $f$ obeys a non linear radial diffusion 
equation in dimensions $d'$ 
\begin{equation}
\frac{\partial f}{\partial t} = \Delta_{d'} f^{\nu}(g,t) \end{equation} where the effective dimension $d'$ is finally given by: \begin{equation} d'-1 = \frac{2(d-1) -\lambda(\nu-1)}{2-\lambda(\nu-1)}
\end{equation}.

Let us first focus on the case $q=\nu=1$, where the diffusion equation is linear. The initial condition
on $x$, $x_0 = 0$, translates into an initial condition on $f(g)$ which 
is a $\delta$-function over the hyper-sphere in dimension $d'=d$, 
of radius $g_0=g(0)$. The solution of the 
radial diffusion equation in $d$ dimension for an isotropic initial condition is obviously constructed as the
superposition of point source solutions of the standard $d$ dimensional diffusion equation (i.e. 
a $d$ dimensional Gaussian), averaged over the position of the starting points, here sitting
on the hyper-sphere ${\cal S}_0$ of radius $g_0$. More explicitly, introducing $d$ dimensional vectors $\vec g$, one
has:
\begin{equation}
f(\vec g) = \int_{{\cal S}_0} \frac{1}{(4 \pi t)^{d/2}} \exp\left(-\frac{(\vec g - \vec g_0)^2}{4 t}\right)
\end{equation}
Introducing the angle $\theta$ between $\vec g$ and $\vec g_0$, one finds:
\begin{equation}
f(\vec g) =  \frac{\Omega_{d-2}}{(4 \pi t)^{d/2}} \exp\left(-\frac{g^2 + g_0^2}{4 t}\right)
\int_0^\pi d\theta \sin^{d-2} \theta \, \, \exp\left(\frac{gg_0 \cos \theta}{2 t}\right).
\end{equation}
Using the following identity: 
\begin{equation}
\int_0^\pi d\theta \sin^{d-2} \theta \, \, \exp\left(\frac{gg_0 \cos \theta}{2 t}\right)= 
\sqrt{\pi} \Gamma(\frac{d-1}{2})
\left(\frac{4t}{gg_0}\right)^{d/2-1} I_{d/2-1}(\frac{gg_0}{2t}),
\end{equation}
where $I$ is the Bessel function, we finally recover the solution of Cox and Ross (see also \cite{Italians}).
% (***see also ITALIANS***). 

The case $q > 1$, $\nu < 1$ leads to the so-called `fast' diffusion equation in $d'$ dimensions. In this
case, an explicit point source solution can be easily constructed for an arbitrary position of the 
point source, and is similar to the $d'=1$, Student-like solution discussed in the main text. Unfortunately,
the general solution for an arbitrary distribution of point sources can no longer be constructed since
the equation is non-linear. The case where these points are on an hyper-sphere is, to the best of our
knowledge, unknown, although approximate solutions could perhaps be constructed both for short and long times.
We leave the investigation of this path for future work.

\section{Appendix C: The SABR Model}

The SABR model \cite{sabr} is a stochastic volatility model of the following form
\begin{eqnarray}
dS &=& S^{\beta} \bar{\sigma} d\omega_1\\
d \bar{\sigma} &=& \lambda \bar{\sigma} d \omega_2
\end{eqnarray}
with
\begin{equation}
<d\omega_1 d\omega_2> = \rho dt
\end{equation}
and
\begin{equation}
\bar{\sigma}(0) = \sigma
\end{equation}
It contains four parameters, $\beta$ and $\rho$ both contribute to the skew ($\beta$ is in fact what we call 
$\alpha$ in the present paper), while $\lambda$ is related to the
curvature of the smile and $\sigma$ is the volatility parameter. Because the log-volatility follows 
a purely diffusive process it can become arbitrarily
large at large times. In our study we assume $\beta$ held fixed, seeing as varying $\rho$ can already change 
the slope of the skew curve.


\begin{thebibliography}{99}
\bibitem{corrqf} Eq. (71) of \cite{qf_borland} is an approximation.
\end{thebibliography}
\end{document}